\documentclass[pdflatex,sn-nature]{sn-jnl}

\usepackage{graphicx}
\usepackage{amsmath,amssymb,amsfonts}
\usepackage{booktabs}
\usepackage{xcolor}

\unnumbered

\begin{document}

\title[Kiloparsec stellar flows]{Contrasting geometries of kiloparsec-scale stellar flows associated with giant Galactic cavities}

\author*[1]{\fnm{Satoya} \sur{Nakano}}\email{nakanosatoya@gmail.com}
\author[2]{\fnm{Kengo} \sur{Tachihara}}

\affil*[1]{\orgname{Independent Researcher}, \country{Japan}}

\affil[2]{\orgdiv{Department of Physics, Graduate School of Science},
\orgname{Nagoya University},
\orgaddress{\street{Furo-cho, Chikusa-ku}, \city{Nagoya}, \state{Aichi},
\postcode{464-8602}, \country{Japan}}}

\abstract{
Giant bubbles and cavities are widespread in star-forming galactic discs, yet the stellar dynamics associated with such structures remain poorly constrained. Here we combine Gaia Data Release 3 with three-dimensional extinction information to identify 3,502 massive-star candidates corresponding to initial masses $M_{\rm ini}\gtrsim8\,M_{\odot}$ within 5~kpc of the Sun. A search of the stellar-density field alone, independent of interstellar-medium morphology and stellar kinematics, reveals two prominent overdensities, R1 and R2, with characteristic scales of approximately 1.3~kpc. Both are subsequently found to be associated with previously identified giant interstellar cavities in projected position and line-of-sight location. After subtracting locally fitted background velocity fields, stars with full three-dimensional velocities reveal coherent divergent in-plane motions across both structures. R1 extends well beyond a known massive-star association and exhibits strongly anisotropic expansion, with one principal strain rate robustly positive and the other consistent with zero. R2 instead exhibits positive strain rates along both principal directions, indicating broadly two-dimensional internal expansion. These results show that density-selected, cavity-associated stellar structures of comparable scale can host coherent kiloparsec-scale flows with contrasting internal flow geometries and need not share a common stellar kinematic organisation.
}
\maketitle

Massive-star feedback reshapes star-forming galactic discs through ionising radiation, stellar winds and supernova explosions, excavating bubbles, shells and cavities in the surrounding interstellar medium \cite{Dale2015}. JWST observations have revealed a highly structured, porous interstellar medium in nearby galaxies \cite{Lee2023,Meidt2023}. In NGC~628 alone, 1,694 bubbles have been identified with radii from a few parsecs to more than 500~pc \cite{Watkins2023}. One of its kiloparsec-scale superbubbles contains a spatially resolved young stellar population together with an expanding multiphase gaseous shell \cite{Mayya2023}. Large superbubble populations are also seen in H\,\textsc{i}, with a recent survey of M31 identifying 118 superbubbles across its disc \cite{Meng2026}. A related configuration is found in the Large Magellanic Cloud, where the approximately 1.4~kpc supergiant shell LMC~4 contains an extended young stellar population \cite{Braun1997}. Together, these observations show that both interstellar structure and recent star formation can remain organised over kiloparsec scales.

Whether the stars associated with such structures retain coherent motions over the same scales is much less clear. An expanding gaseous shell does not itself require an expanding stellar population: once stars form, their trajectories need not follow the subsequent hydrodynamic evolution of the surrounding gas. Gaia studies of nearby OB associations have instead revealed structured velocity fields, including localised expansion, substantial internal substructure and departures from simple monolithic expansion \cite{Kuhn2019,Ward2020,Wright2020}. These studies probe young stellar dynamics on scales of tens to hundreds of parsecs. Beyond individual associations, Gaia-based studies have mapped young and hot stellar populations across several kiloparsecs of the Galactic disc, revealing large-scale stellar structure associated with spiral-arm segments \cite{Zari2021,Poggio2021,GaiaDisc2023}. More recently, SDSS-V spectroscopy combined with Gaia astrometry has revealed large-scale Galactocentric radial-velocity patterns of young OB stars coherent over kiloparsec scales \cite{Zari2025}. For stellar structures associated with giant interstellar cavities, however, the geometry of their internal velocity fields remains poorly constrained.

More direct connections between organised stellar motions and specific Galactic structures have emerged from targeted studies of individual systems. Car~OB1 shows evidence for global expansion on association scales \cite{Melnik2020,Goppl2025}, while coherent outward motions have been reported for groups of O--B2 stars located along the periphery of the Giant Oval Cavity, a structure more than 1~kpc across that spans the Perseus arm \cite{Chen2025}. These studies demonstrate organised stellar motions within OB associations and among stellar groups along the periphery of a kiloparsec-scale cavity. However, such analyses have been centred on previously identified associations or interstellar cavities. It therefore remains unclear whether kiloparsec-scale stellar structures identified from the stellar-density field alone, without reference to interstellar-medium morphology or stellar kinematics, exhibit coherent internal flows, and how their velocity-field geometries compare when analysed within a common framework.

The Milky Way offers the opportunity to address this question through resolved stellar motions, but constructing a uniformly selected sample of massive stars over several kiloparsecs is challenging. Optical samples are strongly affected by spatially variable extinction across the Galactic plane. Spectroscopic catalogues provide secure classifications, but existing Galactic O- and B-star censuses draw on multiple surveys and literature compilations and therefore remain non-uniform in depth and selection \cite{MaizApellaniz2004,Pantaleoni2025}. Gaia Data Release 3 (Gaia DR3) provides homogeneous all-sky astrometry and photometry, together with radial velocities for a subset of stars \cite{GaiaDR3}. Three-dimensional dust reconstructions now map extinction as a function of distance, from the nearby interstellar medium to sightlines extending over several kiloparsecs \cite{Green2019,Vergely2022,Edenhofer2024,Wang2025Dust}. Combining these data enables construction of an extinction-corrected massive-star candidate sample over several kiloparsecs without defining targets from known stellar associations or interstellar structures.

Here we combine Gaia DR3 with three-dimensional extinction information to identify 3,502 massive-star candidates corresponding to initial masses $M_{\rm ini}\gtrsim 8\,M_{\odot}$ within $5~{\rm kpc}$ of the Sun. A search of the stellar-density field alone, without reference to interstellar-medium morphology or stellar kinematics, reveals two prominent overdensities, R1 and R2, each with a characteristic scale of approximately 1.3~kpc. Both are subsequently found to be associated with previously identified giant interstellar cavities in projected position and line-of-sight location \cite{Wang2025Clouds,Vergely2022,Chen2025}. After subtracting locally fitted background velocity fields, stars with full three-dimensional velocity measurements reveal coherent divergent in-plane motions in both structures. R1 extends beyond a previously known massive-star association and exhibits strongly anisotropic expansion, with one principal strain rate robustly positive and the other consistent with zero. R2 instead exhibits positive strain rates along both principal directions, indicating broadly two-dimensional internal expansion. These results show that cavity-associated stellar structures of comparable spatial scale can exhibit coherent kiloparsec-scale flows with contrasting internal flow geometries, and therefore need not share a common stellar kinematic organisation.

\section{Results}

\subsection{Two prominent kiloparsec-scale massive-star overdensities}

\begin{figure}[t]
\centering
\includegraphics[width=\textwidth]{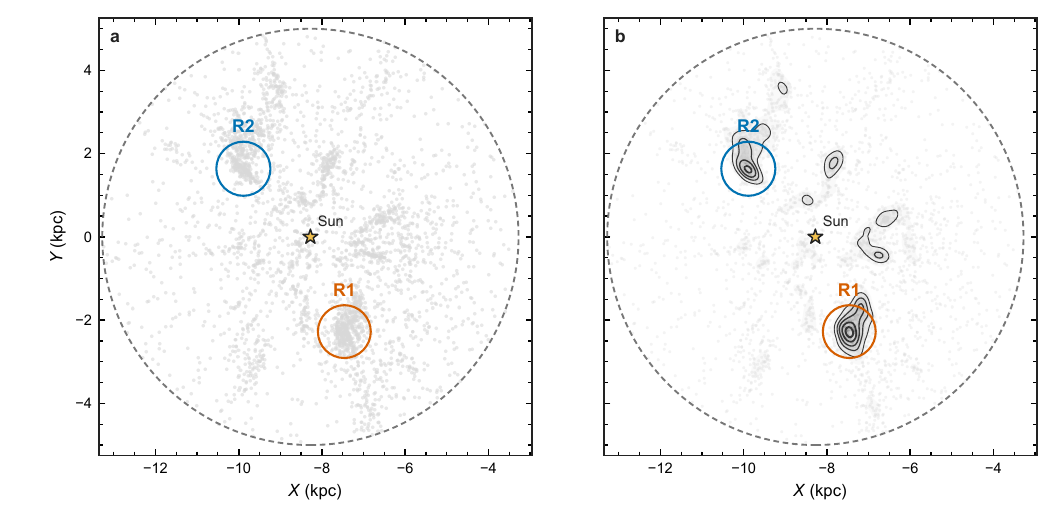}
\caption{\textbf{Two prominent kiloparsec-scale massive-star overdensities within 5~kpc of the Sun.}
\textbf{a,} Distribution of 3,502 massive-star candidates in the Galactic plane. The yellow star marks the Sun, and the dashed grey circle denotes the 5~kpc heliocentric selection boundary. R1 and R2 are shown in orange and blue, respectively. The coloured circles are centred on the stellar-density peaks and have diameters $D_{80}=1.27$~kpc for R1 and $D_{80}=1.30$~kpc for R2.
\textbf{b,} Gaussian-smoothed surface-density distribution of the same stellar sample, using a kernel width of $\sigma=0.12$~kpc. Contours correspond to 0.20, 0.35, 0.50, 0.70 and 0.90 of the maximum smoothed density over the full 5~kpc field. $X$ and $Y$ are Galactocentric Cartesian coordinates in the Galactic plane. The $D_{80}$ circles indicate characteristic spatial scales rather than sharp physical boundaries.}
\label{fig:discovery}
\end{figure}

Our extinction-corrected selection yields 3,502 massive-star candidates corresponding to initial masses $M_{\rm ini}\gtrsim8\,M_{\odot}$ within $5~{\rm kpc}$ of the Sun. Their Galactic-plane distribution is highly structured, with concentrations spanning several hundred parsecs (Fig.~\ref{fig:discovery}a). A search of the stellar surface-density field alone, without reference to known associations, interstellar-medium morphology or stellar motions, identifies R1 and R2 as the two most prominent persistent peaks across multiple smoothing scales (Methods). Figure~\ref{fig:discovery}b shows the stellar surface-density field smoothed with $\sigma=0.12~{\rm kpc}$ for visualisation.

R1 and R2 have adopted centres at $(X,Y)=(-7.462,-2.275)~{\rm kpc}$ and $(-9.887,1.637)~{\rm kpc}$, respectively. Their identification is robust to the adopted smoothing and peak-separation parameters: they remain the two highest distinct peaks across all 15 tested parameter combinations, with peak positions within $0.07~{\rm kpc}$ of the adopted centres (Extended Data Fig.~1). They are also recovered under the Gaia DR3 ESP-HS extinction treatment, and repeating the density search using only the 8,004 ESP-HS-selected candidates absent from the fiducial sample again ranks R1 and R2 as the two highest distinct peaks across all 15 tested parameter combinations (Methods; Extended Data Fig.~2). A targeted test relaxing the initial uncorrected luminosity cut by 3.5~mag recovers only 13 additional candidates around R1 and R2 (Methods). The characteristic size measure $D_{80}$ (Methods) gives $D_{80}=1.27~{\rm kpc}$ for R1 and $D_{80}=1.30~{\rm kpc}$ for R2. Their kiloparsec-scale extents remain robust when photogeometric-distance uncertainties are propagated and when the line-of-sight depth is removed by placing each region at a single region-specific distance (Methods; Extended Data Fig.~3). $D_{80}$ is a characteristic spatial scale rather than a sharp physical boundary.

Because neither stellar kinematics nor interstellar-medium morphology entered the identification of R1 and R2, their stellar kinematics and spatial correspondence with interstellar cavities are post-identification properties rather than selection criteria. We therefore test whether these density-selected structures host coherent large-scale stellar flows.

\subsection{Anisotropic divergent stellar motions in R1}

\begin{figure}[t]
\centering
\includegraphics[width=\textwidth]{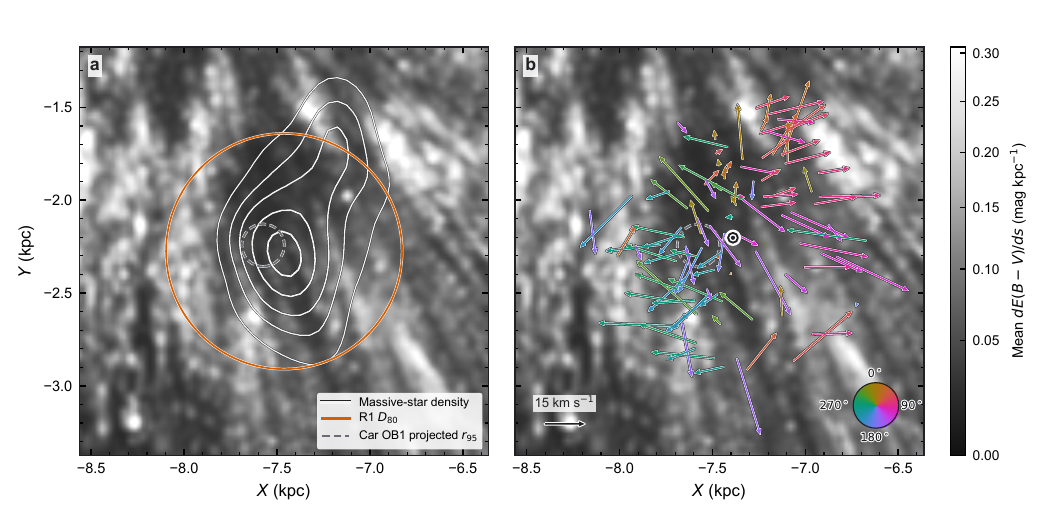}
\caption{\textbf{Kiloparsec-scale structure and residual stellar velocities in R1.}
\textbf{a,} Mean differential reddening, $dE(B-V)/ds$, around R1 from the three-dimensional dust-extinction map \cite{Wang2025Dust}, averaged over $|Z|\leq0.3$~kpc. The low-dust structure coincident with R1 corresponds to the previously recognised Cavity~C in the Galactic molecular-cloud distribution \cite{Wang2025Clouds}. White contours show the massive-star surface-density field at normalised levels of 0.20, 0.35, 0.50, 0.70 and 0.90, using the same global normalisation as in Fig.~\ref{fig:discovery}. The orange circle is centred on the R1 density peak and has a diameter $D_{80}=1.27$~kpc. The grey dashed circle marks the projected 95\%-enclosing radius, $r_{95}=0.115$~kpc, of the published Car~OB1 high-mass-star and OB-candidate sample \cite{Goppl2025}.
\textbf{b,} Residual in-plane velocities of the 106 stars with full three-dimensional velocity measurements within $r\leq0.8~{\rm kpc}$ of the R1 density peak, after subtraction of the locally fitted background velocity field and the remaining bulk motion. Arrow direction and colour encode velocity position angle, and arrow length is proportional to speed. The bullseye marks the stellar centroid used to define radial motions; the reference arrow corresponds to $15\,\mathrm{km\,s^{-1}}$. Vector lengths are capped at $30\,\mathrm{km\,s^{-1}}$ for display. $X$ and $Y$ are Galactocentric Cartesian coordinates in the Galactic plane.}
\label{fig:R1}
\end{figure}

R1 is spatially associated with Cavity~C, a large low-dust structure previously recognised in the Galactic molecular-cloud distribution \cite{Wang2025Clouds} (Fig.~\ref{fig:R1}a). Using the underlying three-dimensional dust-extinction map \cite{Wang2025Dust}, we quantify the three-dimensional correspondence between Cavity~C and the R1 stellar structure. A multiscale analysis detects a persistent dust deficit at the R1 density peak over four consecutive physical smoothing scales, $\sigma=0.10$--$0.20$ kpc. Both the maximum dust-deficit response and its persistence across scales are uncommon among local control positions subjected to the same nine-scale search (Methods; Extended Data Fig.~4). At all four scales, the line-of-sight peak of the dust deficit lies within the central 68\% interval of the stellar $s_{\rm los}$ distribution. The correspondence is therefore supported in both projected position and line-of-sight location and is not tied to a single smoothing choice.

Car~OB1, a previously identified massive-star association in this region \cite{Goppl2025}, occupies only a small part of R1. Assuming a common distance of 2.35~kpc for the published Car~OB1 high-mass-star and OB-candidate sample yields a projected 95\%-enclosing diameter of 0.23~kpc, consistent with the approximately 200~pc extent reported for Car~OB1. R1 therefore traces a stellar structure extending well beyond the spatial footprint of the known association.

We examined the 106 stars with full three-dimensional velocities within 0.8~kpc of the R1 density peak. After subtracting a local affine velocity field fitted to stars in the surrounding annulus at $1.35<r\leq1.75$~kpc and the remaining bulk motion of the R1 sample, the residual field shows coherent outward motions across R1, extending well beyond the spatial footprint of Car~OB1 (Fig.~\ref{fig:R1}b). The flow is strongly direction-dependent rather than consistent with simple isotropic expansion about the stellar centroid. We quantify this geometry below using radial-flow models and the two-dimensional velocity-gradient tensor.

\subsection{Two-dimensional divergent stellar motions in R2}

\begin{figure}[t]
\centering
\includegraphics[width=\textwidth]{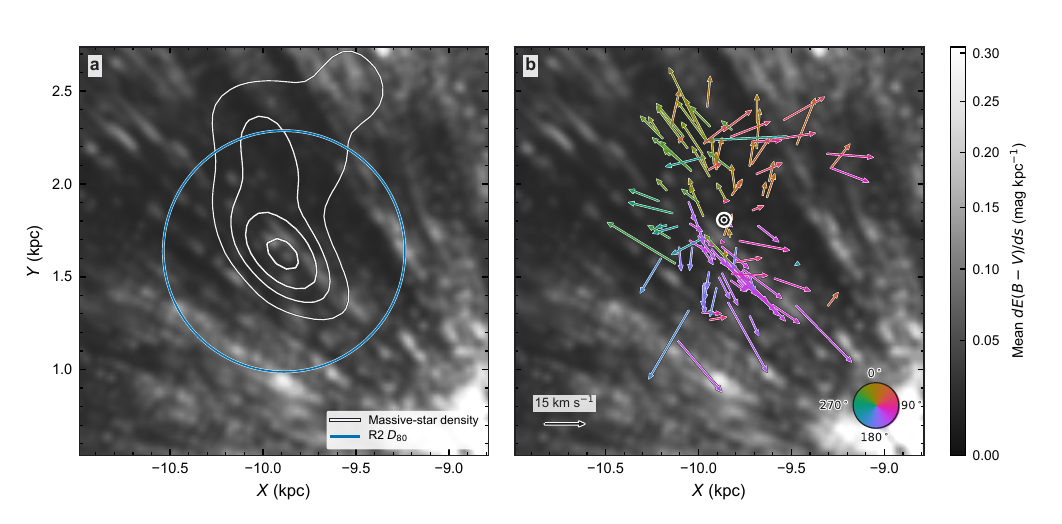}
\caption{\textbf{Kiloparsec-scale structure and residual stellar velocities in R2.}
\textbf{a,} Mean differential reddening, $dE(B-V)/ds$, around R2 from the three-dimensional dust-extinction map \cite{Wang2025Dust}, averaged over $|Z|\leq0.3$~kpc. The low-dust structure coincident with R2 corresponds to the Giant Oval Cavity, previously identified in three-dimensional extinction maps by Vergely et al. \cite{Vergely2022} and subsequently analysed by Chen et al. \cite{Chen2025}. White contours show the massive-star surface-density field at normalised levels of 0.20, 0.35, 0.50, 0.70 and 0.90, using the same global normalisation as in Fig.~\ref{fig:discovery}. The blue circle is centred on the R2 density peak and has a diameter $D_{80}=1.30$~kpc.
\textbf{b,} Residual in-plane velocities of the 118 stars with full three-dimensional velocity measurements within $r\leq0.8~{\rm kpc}$ of the R2 density peak, after subtraction of the locally fitted background velocity field and the remaining bulk motion. Arrow direction and colour encode velocity position angle, and arrow length is proportional to speed. The bullseye marks the stellar centroid used to define radial motions; the reference arrow corresponds to $15\,\mathrm{km\,s^{-1}}$. Vector lengths are capped at $30\,\mathrm{km\,s^{-1}}$ for display. $X$ and $Y$ are Galactocentric Cartesian coordinates in the Galactic plane.}
\label{fig:R2}
\end{figure}

R2 is spatially associated with the Giant Oval Cavity, a large low-extinction structure traced in the three-dimensional extinction maps of Vergely et al. \cite{Vergely2022} and subsequently analysed by Chen et al. \cite{Chen2025} (Fig.~\ref{fig:R2}a). Using the fiducial dust-field construction, we applied the same multiscale analysis used for R1 to the three-dimensional extinction map \cite{Wang2025Dust} and recovered a persistent dust deficit at the R2 density peak over three consecutive physical smoothing scales, $\sigma=0.30$--$0.50~{\rm kpc}$. At the smoothing scale with the strongest projected dust-deficit response, the half-maximum interval of the corresponding line-of-sight deficit profile encompasses the central 68\% interval of the stellar $s_{\rm los}$ distribution (Methods; Extended Data Fig.~4). Its $D_{80}=1.30~{\rm kpc}$ is comparable to the reported extent of the Giant Oval Cavity; Chen et al. instead reported coherent outward motions of O--B2 stellar groups along the cavity periphery \cite{Chen2025}. Because R2 was identified solely from the massive-star density field, without reference to the known cavity, this correspondence was not imposed by the discovery procedure.

We examined the 118 stars with full three-dimensional velocity measurements within $0.80~{\rm kpc}$ of the R2 density peak. This sample has no overlap with the 369-star O--B2 sample analysed by Chen et al. \cite{Chen2025}. After subtracting a local affine velocity field fitted to stars in the surrounding annulus at $1.10<r\leq1.50~{\rm kpc}$ and removing the remaining bulk motion of the R2 sample, the resulting residual velocity field shows coherent outward motions across R2 (Fig.~\ref{fig:R2}b). In contrast to the strongly direction-dependent flow in R1, the residual motions in R2 span a broad range of position angles, forming a broadly two-dimensional divergent pattern. We quantify these flow geometries below using radial-flow models and the two-dimensional velocity-gradient tensor.

\subsection{Contrasting divergent-flow geometries in R1 and R2}

\begin{figure}[t]
\centering
\includegraphics[width=\textwidth]{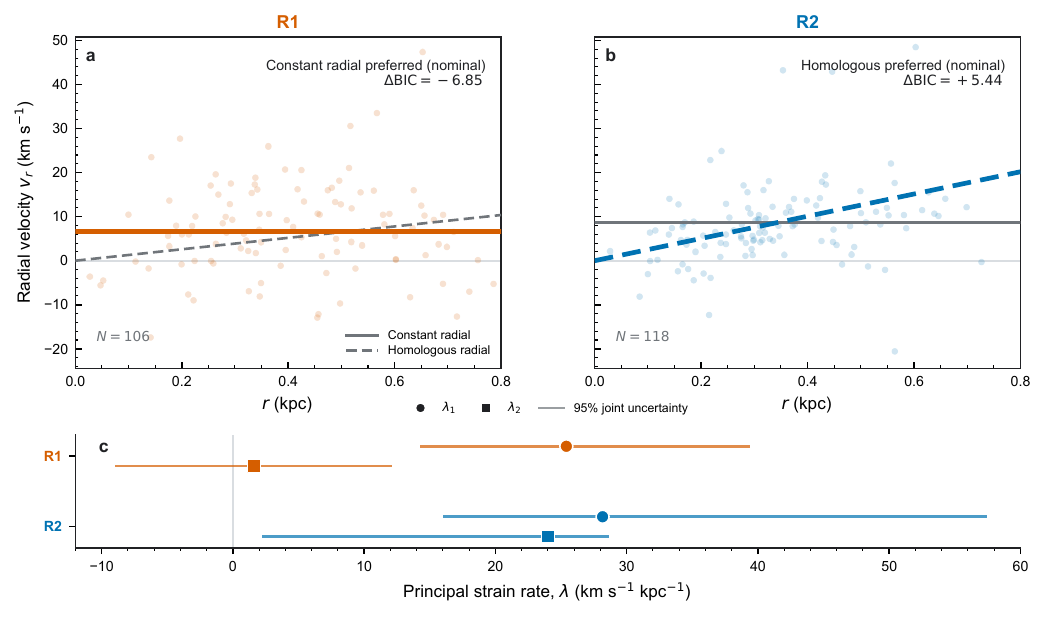}
\caption{\textbf{Radial-flow models and principal strain rates for R1 and R2.}
\textbf{a,b,} Residual radial velocity, $v_r$, as a function of distance from the stellar centroid for stars within $r\leq0.8~{\rm kpc}$ of R1 (\textbf{a}; $n=106$) and R2 (\textbf{b}; $n=118$). Solid lines show constant-speed radial-flow models and dashed lines show homologous models, $v_r=Hr$. The annotations show the nominal model comparison, with $\Delta\mathrm{BIC}\equiv\mathrm{BIC}_{\rm const}-\mathrm{BIC}_{\rm hom}$; negative and positive values favour the constant-speed and homologous models, respectively.
\textbf{c,} Principal strain rates, $\lambda_1$ and $\lambda_2$, given by the eigenvalues of the symmetric component of the two-dimensional velocity-gradient tensor. Circles and squares denote $\lambda_1$ and $\lambda_2$, respectively; horizontal bars show the 95\% joint uncertainty intervals from measurement-error propagation and bootstrap resampling.}
\label{fig:kinematics}
\end{figure}

We first quantified the residual flows with two radial models: a constant-speed model, $v_r=V_{\rm out}$, and a homologous model, $v_r=Hr$, in which radial velocity increases linearly with radius (Fig.~\ref{fig:kinematics}). In R1, 74.5\% of the stars have positive radial velocities, with a median of $6.38\,\mathrm{km\,s^{-1}}$, whereas in R2 the corresponding values are 86.4\% and $7.53\,\mathrm{km\,s^{-1}}$. Using the Bayesian information criterion (BIC), the nominal fits favour the constant-speed model in R1 ($\Delta\mathrm{BIC}\equiv\mathrm{BIC}_{\rm const}-\mathrm{BIC}_{\rm hom}=-6.85$) and the homologous model in R2 ($\Delta\mathrm{BIC}=5.44$). Joint propagation of measurement uncertainties and bootstrap resampling yields median $\Delta\mathrm{BIC}$ values of $-5.37$ and $4.73$, respectively, but the 95\% joint uncertainty intervals include zero in both cases. The radial fits therefore nominally favour different radial scalings in R1 and R2, but neither model preference is robust to measurement and sampling uncertainties.

The measured flow amplitudes imply characteristic kinematic timescales with point estimates ranging from several tens to approximately 100~Myr. Defining $t_{r/V}\equiv(D_{80}/2)/V_{\rm out}$ gives $92.7^{+82.1}_{-29.0}$ Myr for R1 and $78.0^{+54.2}_{-24.8}$ Myr for R2; the corresponding $H^{-1}$ values are $73.2^{+105.3}_{-26.8}$ Myr and $43.5^{+40.5}_{-16.3}$ Myr. The quoted ranges were obtained by inverse-transforming the 95\% joint intervals of the fitted flow amplitudes while holding $D_{80}$ fixed (Methods). These quantities are characteristic flow timescales rather than direct estimates of the ages of R1 and R2.

To characterise the flow geometry without imposing radial symmetry, we fitted a two-dimensional linear velocity field and decomposed the symmetric part of its velocity-gradient tensor into two principal strain rates, $\lambda_1$ and $\lambda_2$. For R1, the nominal values are $\lambda_1=25.4\,\mathrm{km\,s^{-1}\,kpc^{-1}}$ and $\lambda_2=1.6\,\mathrm{km\,s^{-1}\,kpc^{-1}}$. The joint measurement-error and bootstrap analysis gives $\lambda_1=27.2^{+12.1}_{-12.8}\,\mathrm{km\,s^{-1}\,kpc^{-1}}$ and $\lambda_2=1.7^{+10.3}_{-10.6}\,\mathrm{km\,s^{-1}\,kpc^{-1}}$, leaving the former robustly positive while the latter remains consistent with zero. R1 therefore exhibits strongly anisotropic expansion, dominated by a single principal direction.

R2 shows a contrasting tensor structure. Its nominal principal strain rates are $\lambda_1=28.2\,\mathrm{km\,s^{-1}\,kpc^{-1}}$ and $\lambda_2=24.0\,\mathrm{km\,s^{-1}\,kpc^{-1}}$. The joint analysis gives $\lambda_1=31.0^{+26.4}_{-14.9}\,\mathrm{km\,s^{-1}\,kpc^{-1}}$ and $\lambda_2=15.7^{+12.9}_{-13.3}\,\mathrm{km\,s^{-1}\,kpc^{-1}}$, with both remaining positive across their 95\% joint uncertainty intervals. R2 therefore exhibits broadly two-dimensional expansion.

The qualitative contrast in flow geometry persists under reasonable changes in the core radius and background annulus, leave-one-out tests, removal of up to 10\% of the stars with the largest residual speeds, and reweighting for spatial variations in the sampling of the six-dimensional (6D) subset within the fiducial massive-star sample. Across the joint measurement-error and bootstrap realisations, $\lambda_{2,\mathrm{R2}}>\lambda_{2,\mathrm{R1}}$ in 95.2\% of cases, although the 95\% joint uncertainty interval on the direct difference includes zero. Together, these tests consistently recover the same qualitative pattern: strongly anisotropic expansion in R1 and broadly two-dimensional expansion in R2 (Methods; Extended Data Fig.~5).

\section{Discussion}

Previous Gaia studies have revealed expansion and other organised motions within nearby OB associations and young stellar systems \cite{Kuhn2019,Ward2020,Goppl2025}, while coherent outward motions have recently been reported for O--B2 stellar groups along the periphery of the previously identified Giant Oval Cavity \cite{Chen2025}. At larger scales, SDSS-V spectroscopy combined with Gaia astrometry has revealed large-scale Galactocentric radial-velocity patterns of young OB stars coherent over kiloparsec scales across the Galactic disc \cite{Zari2025}. Against this broader context, the key distinction is that R1 and R2 were identified solely from the massive-star density field, without reference to interstellar-medium morphology or stellar kinematics, before their internal residual motions were characterised after local-background subtraction. For R2, this extends the previous picture based on peripheral stellar groups by revealing a kiloparsec-scale massive-star overdensity spatially associated with the low-extinction cavity region and hosting an organised internal divergent flow. More generally, the correspondence of R1 and R2 with giant cavities in projected position and line-of-sight location, together with coherent residual divergence across both structures, shows that density-selected stellar structures associated with giant cavities can host organised internal stellar flows on kiloparsec scales.

The two structures nevertheless exhibit contrasting velocity-field geometries. R1 exhibits strongly anisotropic expansion, with one principal strain rate robustly positive and the other consistent with zero, whereas R2 retains positive strain rates along both principal directions, indicating broadly two-dimensional expansion. This qualitative contrast persists across the robustness tests, including variations in the core and background regions and reweighting for spatial variations in the 6D sampling. Across the joint measurement-error and bootstrap realisations, $\lambda_{2,\mathrm{R2}}>\lambda_{2,\mathrm{R1}}$ in 95.2\% of cases, although the 95\% joint uncertainty interval on the direct difference includes zero. Coherent divergent stellar flows on comparable kiloparsec scales therefore need not share a common internal geometry.

The characteristic kinematic timescales provide an additional constraint on the interpretation of these flows. For R1, $t_{r/V}$ and $H^{-1}$ are approximately 93 and 73 Myr, respectively, whereas R2 gives approximately 78 and 44 Myr. For R2, $t_{r/V}$ is close to the $\sim80$-Myr $R/v$ timescale inferred by Chen et al. \cite{Chen2025} from O--B2 stellar groups along the cavity periphery. This agreement between spatially distinct stellar tracers is also compatible with their interpretation of the Giant Oval Cavity as a long-lived structure maintained by recurrent stellar feedback. Neither $t_{r/V}$ nor $H^{-1}$, however, should be interpreted as a unique dynamical age: the radial-model preference is not robust to measurement and sampling uncertainties, the flows need not be homologous, and anisotropic expansion further limits simple kinematic-age interpretations \cite{Wright2024}. The tens-to-100-Myr characteristic timescales found here are therefore difficult to interpret as the age of a single compact, coeval stellar population undergoing simple ballistic expansion from birth. Any viable interpretation must instead accommodate departures from this simple picture, through non-compact initial conditions, extended or recurrent star formation, multiple stellar generations, time-dependent flows, or a combination of these processes.

The contrasting flow geometries constrain possible origins of the stellar motions but do not uniquely identify them. One possibility is that the stellar motions retain kinematic structure inherited from the natal gas within a hierarchically structured star-forming environment \cite{Ward2020,Wright2020,Lim2019}. This is particularly relevant to R1: recent work on Car~OB1 suggests that its association-scale expansion primarily reflects the initial kinematics of an unbound cloud complex, even though sequential supernova feedback may have shaped later star formation \cite{GopplPreibisch2026}. More generally, anisotropic expansion is observed in Vela~OB2 and direction-dependent expansion appears common among nearby OB associations \cite{Armstrong2022,Quintana2026}. Another, not mutually exclusive possibility is that extended or recurrent star formation and feedback across an evolving complex contribute to the observed coherence \cite{Zucker2022,Mayya2023}. Stellar feedback offers a plausible physical link between the massive stars and the surrounding cavities \cite{Dale2015}. The contrasting R1 and R2 geometries do not, however, require different driving mechanisms. In an inhomogeneous interstellar medium, feedback-driven bubbles can expand preferentially through low-density regions and become highly asymmetric \cite{Fielding2018,ZamoraAviles2019}. Environmental structure can therefore modify cavity evolution even within a common feedback-related framework, although whether such gas asymmetries are reflected in the stellar velocity field remains unclear. The kiloparsec-scale coherence observed here may therefore reflect large-scale organisation of the natal gas, extended or recurrent star formation and feedback, or subsequent dynamical evolution, rather than a single coeval kiloparsec-scale event. The cavity association does not establish that cavity formation directly produced the stellar motions. Any successful scenario must account for coherent divergent stellar motions on kiloparsec scales while accommodating contrasting flow geometries.

The contrasting stellar-flow geometries also bear on the interpretation of large cavities in nearby galaxies. JWST observations reveal that star-forming discs contain numerous bubbles and cavities spanning scales from tens of parsecs to kiloparsecs \cite{Lee2023,Meidt2023,Watkins2023}, and some of the largest structures are associated with extended young stellar populations and expanding multiphase shells \cite{Mayya2023}. Our results show that cavity-associated stellar structures on comparable spatial scales can exhibit coherent divergent motions yet differ in their internal flow geometries. Comparable spatial scales therefore do not imply a common stellar kinematic organisation. Although comparable resolved stellar kinematics remain difficult to obtain in more distant galaxies, combining cavity morphology with stellar kinematics where possible can test whether similar interstellar structures harbour similarly organised stellar flows, providing constraints on the formation and dynamical evolution of cavity--stellar systems that morphology alone cannot provide.

Several limitations preclude direct reconstruction of the dynamical histories of R1 and R2. Full three-dimensional velocities are available for only a subset of the massive-star candidates, although the contrasting flow geometries persist after reweighting for spatial variations in the 6D sampling fraction. Individual stellar ages are insufficiently constrained to determine how the motions developed, while the dynamical relation between the stars and surrounding cavities remains unresolved because stellar motions have not been modelled jointly with gas kinematics and three-dimensional structure. More complete spectroscopy, improved stellar age constraints and spatially resolved H\,\textsc{i}, CO and ionised-gas kinematics will be needed to discriminate among competing evolutionary scenarios. R1 and R2 nevertheless demonstrate that coherent stellar motions can span kiloparsec-scale cavity-associated structures while exhibiting contrasting internal flow geometries. Cavity association and spatial scale alone therefore do not uniquely specify stellar kinematic organisation; resolved stellar kinematics provide an independent dynamical dimension for testing how large-scale interstellar structure relates to the evolution of young stellar populations.

\section{Methods}

\subsection{Stellar distances and three-dimensional extinction correction}

We constructed an initial stellar catalogue by cross-matching Gaia Data Release 3 (Gaia DR3) \cite{GaiaDR3} with the Gaia EDR3 photogeometric-distance catalogue of Bailer-Jones et al. \cite{BailerJones2021} using the Gaia source identifier. We adopted the median of the photogeometric-distance posterior, $r_{\rm med,photogeo}$, as the fiducial stellar distance; $r_{\rm lo,photogeo}$ and $r_{\rm hi,photogeo}$ denote its 16th and 84th percentiles, respectively, and were retained for the uncertainty analyses described below. We restricted the catalogue to sources with $r_{\rm med,photogeo}<5~{\rm kpc}$ and applied the uncorrected absolute-magnitude pre-selection
\begin{equation}
M_{G,\rm obs} \equiv G-5\log_{10}\left(\frac{r_{\rm med,photogeo}}{\rm pc}\right)+5 <0,
\end{equation}
where $G$ is the observed Gaia $G$-band magnitude. This initial luminosity pre-selection was applied before extinction correction and yielded a parent sample of 898,870 sources. No astrometric or radial-velocity quality cuts were imposed at this stage.

We estimated the cumulative reddening to each source using the all-sky three-dimensional dust map of Wang et al. \cite{Wang2025Dust}, accessed through the \texttt{dustmaps3d} package \cite{dustmaps3d2025}. The Galactic longitude, Galactic latitude and adopted photogeometric distance of each source were supplied to the map. For each sightline, the query returned the cumulative $E(B-V)$ to the stellar distance, the local differential reddening $dE(B-V)/ds$, the estimated uncertainty in the cumulative reddening and the maximum reliable distance of the dust reconstruction. We adopted this map as the fiducial extinction reconstruction because it provides these quantities within a single framework for the all-sky, several-kiloparsec volume analysed here, allowing the stellar-extinction correction and the three-dimensional cavity analysis to use a consistent dust reconstruction. Sources entering the fiducial massive-star selection were required to satisfy
\begin{equation}
r_{\rm med,photogeo}\leq d_{\rm max},
\end{equation}
where $d_{\rm max}$ is the maximum reliable distance returned for the corresponding sightline. The differential-reddening information was not required for the stellar extinction correction and was used separately in the three-dimensional cavity analysis described below.

We calculated the bandpass extinctions $A_G$, $A_{\rm BP}$ and $A_{\rm RP}$ with version 1.0 of the \texttt{XP-Extinction-Toolkit} \cite{Zhang2024,ZhangToolkit2024}, using the empirical XP extinction curve implemented in the toolkit and the Gaia DR3 $G$, $G_{\rm BP}$ and $G_{\rm RP}$ passbands. Following Wang et al. \cite{Wang2025Dust}, the cumulative $E(B-V)$ returned by the three-dimensional dust map was supplied directly as $E(440-550)$ to the extinction calculation, without additional rescaling. Because broad-band extinction depends on the stellar spectral energy distribution, we adopted the Gaia DR3 GSP-Phot effective temperature, $T_{\rm eff}$ \cite{Andrae2023}, when a finite estimate was available. The toolkit temperature grid spans $3750\leq T_{\rm eff}\leq19,500~{\rm K}$; finite GSP-Phot temperatures outside this range were evaluated at the nearest grid boundary. For sources without a finite GSP-Phot temperature, we adopted $T_{\rm eff}=6000~{\rm K}$ as a fixed fallback solely for calculating the bandpass-dependent extinction, rather than assuming a hot-star spectral energy distribution a priori. This value was not used as an independent temperature-selection criterion. We assessed the sensitivity of the final massive-star selection to this assumption by repeating the calculation with a nominal hot-star fallback of $T_{\rm eff}=20,000~{\rm K}$, evaluated at the upper toolkit-grid boundary of $19,500~{\rm K}$, as described below. We adopted $\log g=4.5$ for all sources, corresponding to the default value of the toolkit.

The resulting bandpass extinctions were used to derive the intrinsic Gaia colour,
\begin{equation}
(G_{\rm BP}-G_{\rm RP})_0 = (G_{\rm BP}-G_{\rm RP}) - (A_{\rm BP}-A_{\rm RP}),
\end{equation}
and the extinction-corrected absolute $G$-band magnitude,
\begin{equation}
M_{G,0} = G-A_G - 5\log_{10}\left(\frac{r_{\rm med,photogeo}}{10~{\rm pc}}\right).
\end{equation}
These extinction-corrected quantities were used to construct the fiducial massive-star sample described in the following subsection.

\subsection{Massive-star sample selection}

We selected massive-star candidates in the extinction-corrected Gaia colour--absolute-magnitude diagram, following the photometric selection approach of Nakano et al. \cite{Nakano2025}. The selection boundary was constructed from PARSEC v1.2S stellar-evolution models \cite{Bressan2012,Chen2015} with $Z_{\rm ini}=0.0152$, corresponding to the solar metallicity adopted by the PARSEC models, and was designed to encompass the evolutionary loci of stars with initial masses $M_{\rm ini}\gtrsim8~M_{\odot}$, including both blue and evolved massive-star phases.

We combined PARSEC loci tracing the faint boundary of the massive-star region in intrinsic $(G_{\rm BP}-G_{\rm RP})_0$ versus $M_{G,0}$ space. Where multiple model points shared the same intrinsic colour, we retained the faintest point, that is, the point with the largest $M_G$. The resulting 109 boundary points were ordered by intrinsic colour and linearly interpolated to define a continuous selection boundary,
\begin{equation}
M_{G,\rm boundary} = f\!\left[(G_{\rm BP}-G_{\rm RP})_0\right],
\end{equation}
over the colour interval
\begin{equation}
-0.506\leq(G_{\rm BP}-G_{\rm RP})_0\leq3.331.
\end{equation}
No extrapolation was performed outside this interval. The adopted boundary and the extinction-corrected colour--magnitude distribution are shown in Supplementary Fig.~1; the boundary points are provided with the accompanying data and code.

A source was classified as a massive-star candidate when its intrinsic colour lay within this interval and
\begin{equation}
M_{G,0}\leq M_{G,\rm boundary}.
\end{equation}
Before applying the dust-map reliability requirement, this photometric criterion selected 3,566 sources from the parent sample. Requiring each source to lie within the reliable distance range of the Wang et al. three-dimensional extinction map, as described above, removed 64 sources and yielded the fiducial spatial sample of 3,502 massive-star candidates used for the spatial analyses throughout this work. No additional cut on effective temperature, radial velocity or astrometric quality was imposed in defining this sample.

\subsection{Validation and robustness of the massive-star selection}

We first assessed the sensitivity of the photometric selection to the fallback temperature adopted for sources without a finite GSP-Phot temperature. Repeating the extinction correction with a nominal hot-star fallback of $T_{\rm eff}=20,000~{\rm K}$ instead of the fiducial $6000~{\rm K}$ increased the number of sources satisfying the final massive-star selection from 3,502 to 4,628. The fiducial $6000~{\rm K}$ assumption is therefore conservative with respect to sample inclusion relative to a hot-star fallback.

Because the parent sample was defined using the uncorrected luminosity cut $M_{G,\rm obs}<0$, we separately tested whether spatially varying extinction could preferentially remove massive stars from the surroundings of R1 and R2. We examined sources within a projected Galactic-plane distance of $1.4~{\rm kpc}$ from either fiducial centre, extending the initial luminosity range to $0\leq M_{G,\rm obs}<3.5$. Sightlines capable of producing sufficient extinction to move a source across the adopted PARSEC boundary were identified using the Wang et al. three-dimensional reddening map \cite{Wang2025Dust}, after which the corresponding Gaia sources were evaluated at their individual photogeometric distances using the same dust-map reliability criterion, XP-based extinction correction and colour--magnitude selection as for the fiducial sample. Only 13 additional massive-star candidates were recovered: six around R1 and seven around R2, all with $0\leq M_{G,\rm obs}<1$. No candidates were recovered at $1\leq M_{G,\rm obs}<3.5$. Repeating the sightline search at finer angular sampling across the full R1 and R2 footprints identified no further candidates. Within the tested R1/R2 regions and luminosity range, the initial uncorrected luminosity cut therefore has only a minor effect on the local massive-star census and is unlikely to account for the observed overdensities through differential extinction.

We next evaluated the fiducial selection against two external spectroscopic catalogues. For the Galactic O-Star Catalog (GOSC) \cite{MaizApellaniz2004,MaizApellaniz2013}, we linked catalogue entries to Gaia DR3 through their 2MASS identifiers using the \texttt{gaiadr3.tmass\_psc\_xsc\_best\_neighbour} table. A GOSC entry was counted as recovered if any of its matched Gaia sources satisfied the fiducial photometric selection. Among the 516 GOSC entries with at least one matched Gaia DR3 source in the parent sample, 228 were recovered, corresponding to a conditional recovery fraction of 44.2\%. This fraction does not quantify losses introduced by the initial luminosity pre-selection used to construct the parent sample. For reference, a simple selection based only on $T_{\rm eff,GSP\mbox{-}Phot}>10,000~{\rm K}$ selected 27,543 sources from the same parent sample and recovered 191 of the 516 GOSC entries (37.0\%).

We also cross-matched the fiducial sample with the ALS~III catalogue \cite{Pantaleoni2025} using Gaia DR3 source identifiers, restricting the comparison to its G3 subset. Of 2,158 G3 sources present in the parent sample, 757 (35.1\%) were recovered. Restricting the comparison further to 1,496 sources with secure O--B2 spectral classifications yielded 584 recovered stars (39.0\%). The subtype-specific recovery fractions were 37.9\% for O stars, 44.0\% for B0 stars, 42.9\% for B1 stars and 30.8\% for B2 stars. These comparisons provide external spectroscopic recovery benchmarks rather than estimates of the completeness or purity of the photometrically selected sample, because the reference catalogues are not volume-complete and have heterogeneous selection functions. Detailed counts are given in Supplementary Table~1.

Finally, we tested the dependence of the spatial results on the adopted extinction treatment by constructing an alternative sample using Gaia DR3 ESP-HS extinction estimates from the hot-star Apsis analysis \cite{Creevey2023,Fouesneau2023}. For sources with finite, non-negative $A_{G,\rm ESP\mbox{-}HS}$ and $E(G_{\rm BP}-G_{\rm RP})_{\rm ESP\mbox{-}HS}$ estimates, we recomputed the intrinsic colour and absolute magnitude using these quantities and applied the same colour range and PARSEC-based selection boundary as in the fiducial analysis. Sources without usable ESP-HS extinction estimates retained their fiducial classification. To keep the spatial domain fixed, we retained the same dust-map distance-reliability requirement in both selections. The resulting hybrid sample contained 11,101 candidates, of which 3,097 were shared with the fiducial 3,502-source sample, 8,004 were selected only in the alternative sample and 405 only in the fiducial sample.

To compare the smoothed density morphology of the two samples, we constructed the fiducial and alternative surface-density fields on the same $0.05~{\rm kpc}$ grid and smoothed both with a Gaussian kernel of $\sigma=0.30~{\rm kpc}$. We evaluated the correlation over grid cells within $5~{\rm kpc}$ of the Sun for which the smoothed density was non-zero in at least one of the two fields. The fields were strongly correlated, with Pearson and Spearman correlation coefficients of 0.945 and 0.929, respectively, and both R1 and R2 were recovered in the alternative sample (Extended Data Fig.~2a,b).

As a more stringent test, we repeated the density analysis using only the 8,004 ESP-HS-selected sources that were absent from the fiducial sample. R2 and R1 ranked first and second, respectively, among the distinct density peaks across all 15 combinations of Gaussian smoothing scale and minimum peak separation used in the identification-robustness tests. The recovered R1 position remained within $0.09~{\rm kpc}$ of the fiducial centre across all settings, whereas R2 remained within $0.10~{\rm kpc}$ for $\sigma\leq0.15~{\rm kpc}$ and shifted by $0.34~{\rm kpc}$ at $\sigma=0.20~{\rm kpc}$. For a given smoothing scale, the recovered peak positions were unchanged across the three tested minimum peak separations. The identification of R1 and R2 is therefore not specific to the fiducial extinction treatment or to the stars selected by it, although the precise R2 peak position becomes sensitive to stronger smoothing in the alternative-only sample (Extended Data Fig.~2c,d).

\subsection{Galactocentric coordinates and stellar velocities}

We transformed the stellar positions to a right-handed Galactocentric Cartesian coordinate system using the adopted photogeometric distance $d=r_{\rm med,photogeo}$. We adopted a Galactocentric distance of the Sun of $R_0=8.277~{\rm kpc}$ \cite{Gravity2022} and placed the Sun at $(X,Y,Z)=(-R_0,0,0)$. The positive $X$-direction points from the Sun towards the Galactic centre, the $Y$-axis points in the direction of Galactic rotation and the $Z$-axis towards the north Galactic pole. For Galactic longitude $l$ and latitude $b$, the coordinates were calculated as
\begin{align}
X &= d\cos b\cos l-R_0,\\
Y &= d\cos b\sin l,\\
Z &= d\sin b.
\end{align}
The spatial analyses use the projected Galactocentric coordinates $(X,Y)$.

For stars with the required kinematic measurements, we transformed the Gaia DR3 equatorial coordinates, proper motions and radial velocities to heliocentric Galactic Cartesian velocities using \texttt{Astropy} \cite{Astropy2022}. The heliocentric velocity axes were defined with $v_{x,\rm hel}$ positive towards the Galactic centre, $v_{y,\rm hel}$ in the direction of Galactic rotation and $v_{z,\rm hel}$ towards the north Galactic pole. We converted these velocities to the Galactocentric rest frame by adding the adopted total velocity of the Sun with respect to the Galactic centre,
\begin{equation}
(U_\odot,V_\odot,W_\odot) = (9.3,251.5,8.59)~{\rm km\,s^{-1}},
\end{equation}
following the Galactic reference frame adopted in Gaia DR3 disc kinematic analyses \cite{GaiaDisc2023}. Thus,
\begin{equation}
(V_X,V_Y,V_Z) = (v_{x,\rm hel},v_{y,\rm hel},v_{z,\rm hel}) + (U_\odot,V_\odot,W_\odot).
\end{equation}

Before modelling the local velocity fields, we removed a reference axisymmetric Galactic rotation field. We adopted a constant circular speed of $V_{\rm circ}=236~{\rm km\,s^{-1}}$, consistent with the circular speed at the Solar radius measured by Reid et al. \cite{Reid2019}. At the Galactocentric cylindrical radius
\begin{equation}
R=(X^2+Y^2)^{1/2},
\end{equation}
the corresponding circular-velocity components in our Cartesian convention are
\begin{align}
V_{X,\rm circ} &= V_{\rm circ}\frac{Y}{R},\\
V_{Y,\rm circ} &= -V_{\rm circ}\frac{X}{R}.
\end{align}
We then defined the in-plane velocities relative to this circular reference field as
\begin{align}
u &= V_X-V_{X,\rm circ},\\
v &= V_Y-V_{Y,\rm circ}.
\end{align}
These axisymmetric-rotation-corrected velocities were used as the input to the local background-velocity modelling described below. The selection and quality requirements for stars with full six-dimensional phase-space information are described separately in the subsection ``Six-dimensional kinematic sample and spatial sampling''.

\subsection{Stellar surface-density field and identification of R1 and R2}

We searched for large-scale stellar overdensities using only the projected distribution of the 3,502 stars in the fiducial massive-star sample, without reference to interstellar-medium morphology or stellar kinematics. Stellar positions were binned in the Galactocentric $X$--$Y$ plane on a regular Cartesian grid with a spacing of $0.05~{\rm kpc}$. The initial search grid spanned $-12.8\leq X\leq-4.5~{\rm kpc}$ and $-5.2\leq Y\leq5.2~{\rm kpc}$. The initial search was performed over this rectangular domain without an additional mask for the heliocentric $5~{\rm kpc}$ selection footprint.

The initial identification was performed using a multi-scale peak search. We constructed smoothed density maps using Gaussian filtering implemented in \texttt{SciPy} \cite{Virtanen2020}, with Gaussian widths of $\sigma=0.10$, 0.15 and $0.20~{\rm kpc}$. At each smoothing scale, local density maxima were identified using a minimum separation of $0.40~{\rm kpc}$. For each smoothed map, we calculated the median density and the robust dispersion, $\sigma_{\rm rob}=1.4826\,{\rm MAD}$, where MAD is the median absolute deviation over the rectangular search domain. A local maximum was retained when its density exceeded the median by at least $3\sigma_{\rm rob}$. We defined the corresponding robust density-excess statistic as
\begin{equation}
s_i= \frac{ \Sigma_i(X_i,Y_i)-{\rm med}[\Sigma_i] }{ \sigma_{{\rm rob},i} },
\end{equation}
where $\Sigma_i(X,Y)$ is the smoothed density field at smoothing scale $i$, $(X_i,Y_i)$ is the position of the detected maximum, ${\rm med}[\Sigma_i]$ is the median density over the rectangular search domain and $\sigma_{{\rm rob},i}$ is the corresponding robust dispersion. This statistic was used as a relative measure of peak prominence and was not interpreted as a formal Gaussian significance. Detections obtained at different smoothing scales were associated when separated by no more than $0.30~{\rm kpc}$, and only peaks detected at two or more of the three smoothing scales were retained as persistent stellar-density peaks.

This procedure yielded 17 persistent peaks. We ranked these peaks by their maximum robust density excess, $s_{\max}\equiv\max_i(s_i)$. The two highest values were 49.96 and 40.56, compared with 20.92 for the third-ranked peak. We therefore selected the two highest-ranked persistent overdensities for the subsequent analysis and denote them R1 and R2.

For each persistent peak, the adopted position was calculated as the density-excess-weighted mean of its positions in the individual smoothed maps:
\begin{align}
X_{\rm peak} &= \frac{\sum_i s_i X_i}{\sum_i s_i},\\
Y_{\rm peak} &= \frac{\sum_i s_i Y_i}{\sum_i s_i},
\end{align}
where the sums extend over the smoothing scales at which the peak was detected. The adopted analysis centres of R1 and R2 are
\begin{align}
(X_{\rm R1},Y_{\rm R1}) &= (-7.462,-2.275)~{\rm kpc},\\
(X_{\rm R2},Y_{\rm R2}) &= (-9.887,1.637)~{\rm kpc}.
\end{align}
These coordinates are density-excess-weighted centres of the multi-scale peak detections rather than sub-pixel refinements of a single density map. They were held fixed in the subsequent spatial and kinematic analyses.

For Fig.~1, we constructed the displayed density field on a $0.05~{\rm kpc}$ grid spanning $-13.35\leq X\leq-2.95~{\rm kpc}$ and $-5.25\leq Y\leq5.25~{\rm kpc}$ and adopted a Gaussian smoothing width of $\sigma=0.12~{\rm kpc}$. The smoothed field was normalised by its maximum over the displayed grid. This smoothing scale was used for visualisation and was distinct from the multi-scale procedure used to determine the adopted centres.

We further tested the robustness of the R1 and R2 identification against both the smoothing scale and the minimum separation imposed between distinct local maxima. Using the same $0.05~{\rm kpc}$ grid as for Fig.~1, we repeated the peak search for Gaussian widths of $\sigma=0.08$, 0.10, 0.12, 0.15 and $0.20~{\rm kpc}$ and minimum peak separations of 0.40, 0.60 and $0.80~{\rm kpc}$, giving 15 parameter combinations. For each smoothing scale, the smoothed density field was computed over the full rectangular grid, after which grid cells outside the actual heliocentric $5~{\rm kpc}$ selection footprint were masked before identifying local maxima. At each setting, pixel-scale local maxima were ranked by their smoothed density, and lower-ranked maxima lying within the specified minimum separation of a higher-ranked peak were discarded. To associate detections of the same structures across the parameter grid, the highest-ranked detected peak within $0.50~{\rm kpc}$ of each adopted centre was matched to R1 or R2. This matching radius was used only to track the structures between robustness tests and did not enter their initial identification or define a physical aperture. With this additional footprint restriction, R1 and R2 remained the two highest-ranked distinct stellar-density peaks across all 15 tested combinations, and the corresponding peak positions remained within $0.07~{\rm kpc}$ of their adopted centres (Extended Data Fig.~1).

\subsection{Characteristic spatial scales and spatial robustness}

We quantified the characteristic projected size of each overdensity from its background-subtracted radial stellar-density profile. Using the adopted R1 and R2 centres defined above, we counted stars in concentric annuli of width $0.05~{\rm kpc}$ out to a radius of $2.0~{\rm kpc}$. To account for the local stellar background, we repeated the measurement using six background annuli: $0.9<r\leq1.3$, $1.0<r\leq1.4$, $1.1<r\leq1.5$, $1.2<r\leq1.6$, $1.3<r\leq1.7$ and $1.4<r\leq1.8~{\rm kpc}$. For a background annulus containing $N_{\rm bg}$ stars over area $A_{\rm bg}$, we estimated the background surface density as $\Sigma_{\rm bg}=N_{\rm bg}/A_{\rm bg}$. For each radial annulus $j$, with observed count $N_j$ and area $A_j$, the background-subtracted excess was
\begin{equation}
\Delta N_j=N_j-\Sigma_{\rm bg}A_j.
\end{equation}
For the purpose of locating the radial extent of the excess, we adopted the Poisson-based uncertainty estimate
\begin{equation}
\epsilon_j=\left[N_j+\left(\frac{A_j}{A_{\rm bg}}\right)^2N_{\rm bg}\right]^{1/2}.
\end{equation}
For radial bins that overlap the background annulus, $N_j$ and $N_{\rm bg}$ are not statistically independent; the covariance introduced by this overlap is not included in the expression above. We therefore used $\epsilon_j$ only as a Poisson-based scale for defining the radial edge, rather than as a formal measurement uncertainty.

We defined $z_j=\Delta N_j/\epsilon_j$ and used this quantity to determine the outer radius over which the stellar excess was integrated. Starting from the centre and proceeding outwards, once at least one radial bin at $r>0.15~{\rm kpc}$ had $z_j>1$, we continued outwards and defined $r_{\rm edge}$ as the inner boundary of the first sequence of four consecutive bins with $z_j\leq1$. If no such sequence was found before $2.0~{\rm kpc}$, we set $r_{\rm edge}=2.0~{\rm kpc}$. We then formed the cumulative background-subtracted excess using only radial bins whose centres satisfied $r_{c,j}<r_{\rm edge}$. Because individual background-subtracted bins can fluctuate downwards, we replaced the cumulative profile by its running maximum to obtain a non-decreasing profile. We defined $r_{80}$ as the radius at which this profile reached 80\% of its final cumulative value over the included bins, using linear interpolation between radial-bin centres, and defined
\begin{equation}
D_{80}=2r_{80}.
\end{equation}
For each region, the adopted $D_{80}$ was the median of the six values obtained using the different spatial-background annuli. This gives $D_{80}=1.27~{\rm kpc}$ for R1 and $D_{80}=1.30~{\rm kpc}$ for R2. These values are used as characteristic spatial scales and do not represent sharp physical boundaries. The background annuli used here serve only to estimate the local stellar-density background and are distinct from the kinematic background annuli used below.

We assessed the sensitivity of $D_{80}$ to photogeometric-distance uncertainties using 1,000 Monte Carlo realisations. For each star, we used the median, 16th-percentile and 84th-percentile photogeometric distances from Bailer-Jones et al. \cite{BailerJones2021}, denoted $d_{\rm med}$, $d_{\rm lo}$ and $d_{\rm hi}$, and defined asymmetric dispersions in logarithmic distance as
\begin{equation}
\sigma_-=\ln d_{\rm med}-\ln d_{\rm lo},\qquad
\sigma_+=\ln d_{\rm hi}-\ln d_{\rm med}.
\end{equation}
In each realisation, we drew an independent standard-normal variate $q$ for every star and assigned
\begin{equation}
d=
\begin{cases}
\exp(\ln d_{\rm med}+q\sigma_-), & q<0,\\
\exp(\ln d_{\rm med}+q\sigma_+), & q\geq0.
\end{cases}
\end{equation}
The Galactocentric $X$ and $Y$ coordinates were then recomputed, and the complete $D_{80}$ measurement about the fixed adopted R1 and R2 centres, including all six spatial-background choices, was repeated for each realisation.

As a limiting-case test of whether the projected extents could be produced by line-of-sight distance structure, we also constructed depth-collapsed versions of R1 and R2. For each region, we adopted a single region-specific heliocentric distance equal to the median photogeometric distance of stars lying within the fiducial radius $D_{80}/2$ of the adopted centre. We then placed the full spatial sample at that region-specific distance while preserving the sky position of every star, thereby removing the individual line-of-sight depth. In the resulting projected distribution, an initial position was estimated from the median projected coordinates of the depth-collapsed stars whose fiducial positions lay within $0.30~{\rm kpc}$ of the adopted centre. We constructed a local density map on a $0.05~{\rm kpc}$ grid using Gaussian smoothing with $\sigma=0.12~{\rm kpc}$ and reidentified the local density peak as the maximum of the smoothed density field within $0.40~{\rm kpc}$ of this initial position. We then remeasured $D_{80}$ about the reidentified peak using the same radial-profile estimator and the same six spatial-background annuli as in the fiducial analysis. The dependence of the characteristic spatial scales on background choice, photogeometric-distance uncertainty and removal of the line-of-sight depth is summarised in Extended Data Fig.~3.

For the projected comparison with Car~OB1 shown in Fig.~\ref{fig:R1}, we used the published high-mass-star and OB-candidate sample \cite{Goppl2025}. All catalogue entries with finite sky coordinates were placed at a common heliocentric distance of $2.35~{\rm kpc}$ and transformed to the same Galactocentric $X$--$Y$ coordinates used for R1. We defined the projected Car~OB1 centre as the median $X$ and $Y$ coordinates of this sample and defined $r_{95}$ as the 95th percentile of the projected distances from that centre. This gives $r_{95}=0.115~{\rm kpc}$, corresponding to a 95\%-enclosing diameter of $0.229~{\rm kpc}$.

\subsection{Three-dimensional association with interstellar cavities}

We quantified the correspondence between the density-selected stellar structures and the surrounding interstellar cavities using the three-dimensional dust-extinction map of Wang et al. \cite{Wang2025Dust}. The association was evaluated only after the R1 and R2 stellar-density peaks had been defined, and neither the positions nor the shapes of known cavities were used to determine the stellar structures. R1 spatially overlaps the region associated with the previously identified Cavity~C \cite{Wang2025Clouds}, whereas R2 spatially overlaps the Giant Oval Cavity previously identified in three-dimensional extinction maps by Vergely et al. \cite{Vergely2022} and subsequently analysed by Chen et al. \cite{Chen2025}. These literature identifications provide the physical context for the low-dust structures, but the quantitative projected dust-deficit analysis described below was anchored to the fixed R1 and R2 stellar-density peaks and did not require a prescribed cavity boundary, ellipse, position angle or fitted cavity centre. Throughout this work, ``cavity-associated'' denotes the combination of projected correspondence, a persistent local dust deficit and line-of-sight consistency; it does not imply a uniquely defined cavity boundary or a causal origin for the stellar structure.

We sampled the Wang et al. differential-reddening field, $dE(B-V)/ds$, on local Galactocentric $X$--$Y$ grids with a spacing of $0.04~{\rm kpc}$. Here, $s$ in $dE(B-V)/ds$ denotes the path-length variable of the differential-reddening quantity; the local line-of-sight coordinate introduced below is denoted $s_{\rm los}$. To construct the fiducial in-plane dust field while reducing sensitivity to any single vertical slice, we averaged the map over $|Z|\leq0.30~{\rm kpc}$ using vertical layers spaced by $0.05~{\rm kpc}$. At each $X$--$Y$ position, the average was retained only when at least 50\% of the sampled vertical layers contained valid dust-map values. We searched for extended local dust deficits using a multiscale Difference-of-Gaussians (DoG) statistic. For an inner Gaussian smoothing scale $\sigma$, we defined
\begin{equation}
C_\sigma(X,Y)=G_{2.5\sigma}[D(X,Y)]-G_{\sigma}[D(X,Y)],
\end{equation}
where $D(X,Y)$ is the vertically averaged differential-reddening field and $G_\sigma$ denotes Gaussian smoothing with width $\sigma$. Positive $C_\sigma$ therefore corresponds to a region in which the differential reddening on scale $\sigma$ is lower than in its broader surroundings. We evaluated nine pre-defined inner smoothing scales, $\sigma=0.08$, 0.10, 0.12, 0.16, 0.20, 0.25, 0.30, 0.40 and $0.50~{\rm kpc}$; the corresponding outer smoothing scale was $2.5\sigma$.

At each smoothing scale, the observed DoG contrast was evaluated at the fixed stellar-density peak of R1 or R2. We compared this value with DoG contrasts at local control positions in the annulus $0.80\leq r\leq2.00~{\rm kpc}$ from the same stellar-density peak. For each scale, we defined the robust contrast statistic
\begin{equation}
z_\sigma=\frac{C_\sigma-C_{{\rm med},\sigma}}{1.4826\,{\rm MAD}_\sigma},
\end{equation}
where $C_{{\rm med},\sigma}$ and ${\rm MAD}_\sigma$ are the median and median absolute deviation of the DoG contrasts at the valid control positions. This statistic measures the prominence of the local dust deficit relative to its surroundings and was not interpreted as a formal Gaussian significance.

Because nine smoothing scales were examined, we accounted empirically for the search over scale by applying the same multiscale statistic to control positions that were valid across all nine smoothing scales. We defined $T=\max_\sigma z_\sigma$ and compared the observed value, $T_{\rm obs}$, with the distribution of $T$ obtained from these local controls. The max-response tail fraction was defined as the fraction of control positions with $T\geq T_{\rm obs}$. We separately quantified persistence across scale by recording the longest sequence of consecutive smoothing scales for which $z_\sigma>2$ and comparing the observed sequence length with the corresponding distribution among the control positions. The resulting persistence tail fraction is the fraction of controls with an equally long or longer sequence. These quantities are empirical spatial tail fractions and are not formal Gaussian $p$-values. For the fiducial $|Z|\leq0.30~{\rm kpc}$ field, R1 shows $z_\sigma>2$ over four consecutive scales from $\sigma=0.10$ to $0.20~{\rm kpc}$, with a maximum at $\sigma=0.16~{\rm kpc}$; its max-response and persistence tail fractions are 3.7\% and 4.8\%, respectively. R2 shows $z_\sigma>2$ over three consecutive scales from $\sigma=0.30$ to $0.50~{\rm kpc}$, with a maximum at $\sigma=0.30~{\rm kpc}$; the corresponding tail fractions are 9.6\% and 7.7\%.

We further tested whether the projected dust deficits occur at distances consistent with the stellar structures. For each region, we defined an in-plane heliocentric line-of-sight coordinate $s_{\rm los}$ directed from the Sun towards the adopted stellar-density peak and a perpendicular in-plane coordinate $t$. We selected the massive-star candidates lying within the fiducial radius $D_{80}/2$ in the Galactocentric $X$--$Y$ plane and calculated the 16th, 50th and 84th percentiles of their $s_{\rm los}$ distribution. For each DoG smoothing scale, we then sampled the corresponding DoG contrast field along $s_{\rm los}$ at the transverse coordinates $t$ of these stars and averaged the resulting values at each $s_{\rm los}$, requiring valid map values for at least 50\% of the sampled stellar transverse coordinates. This construction places the stars and the DoG contrast in the same line-of-sight coordinate and avoids comparing the stellar photogeometric distance directly with a differently defined projected coordinate.

For each line-of-sight DoG-contrast profile, we identified its maximum and the contiguous interval around that maximum for which the positive DoG contrast remained above half of the peak value. For R1, all four smoothing scales with $z_\sigma>2$ place the DoG-contrast maximum within the 16th--84th percentile interval of the stellar $s_{\rm los}$ distribution. For R2, at the scale of maximum projected DoG response, $\sigma=0.30~{\rm kpc}$, the half-maximum interval of the line-of-sight DoG-contrast profile encompasses the stellar 16th--84th percentile interval. The smoothing scale that maximised the projected DoG response was used only to summarise and display the line-of-sight correspondence; the prominence of the projected dust deficit was assessed independently through the full nine-scale control analysis described above.

Finally, we tested the sensitivity of the cavity-association statistics to the adopted vertical averaging range and local-control region. We repeated the nine-scale analysis for dust fields averaged over $|Z|\leq0.10$, 0.20 and $0.30~{\rm kpc}$ and for control annuli of $0.70$--$1.80$, $0.80$--$2.00$ and $0.90$--$2.20~{\rm kpc}$, giving nine combinations for each region. These tests preserved the fixed stellar-density centres and varied only the construction of the dust field and its local comparison region.

For R1, the max-response tail fraction ranged from 3.4\% to 8.1\% and the persistence tail fraction from 1.8\% to 7.8\%, with the scale of maximum response at either $\sigma=0.12$ or $0.16~{\rm kpc}$. For R2, the scale of maximum response remained $\sigma=0.30~{\rm kpc}$ in all nine combinations, but its multiscale persistence was more sensitive to the adopted choices: the max-response tail fraction ranged from 8.9\% to 18.1\%, while the longest run of consecutive scales with $z_\sigma>2$ ranged from zero to three, corresponding to persistence tail fractions of 7.6\%--100\%. Extended Data Fig.~4 shows the fiducial multiscale control comparison for $|Z|\leq0.30~{\rm kpc}$ and a control annulus of $0.80$--$2.00~{\rm kpc}$, together with the corresponding line-of-sight stellar--dust comparison.

\subsection{Six-dimensional kinematic sample and spatial sampling}

For the kinematic analysis, we selected the subset of the fiducial massive-star sample with the Gaia DR3 astrometric and radial-velocity measurements required to construct full three-dimensional space motions \cite{Lindegren2021,Katz2023}. We required a renormalised unit weight error (RUWE) below 1.4, a radial-velocity uncertainty below $10~{\rm km\,s^{-1}}$ and at least five radial-velocity transits. These criteria yielded 846 stars with full six-dimensional phase-space information. The positions and velocities of these stars were calculated using the same photogeometric distances and Galactocentric transformation described above. Throughout this subsection, the term 6D sampling fraction refers to the fraction of the 3,502-star fiducial massive-star sample represented in this kinematic subset and should not be interpreted as the completeness of the massive-star catalogue relative to the underlying Galactic population.

Because Gaia radial velocities are available for only a subset of the fiducial sample, we quantified the spatial sampling of the 6D subset within R1 and R2 before analysing their internal velocity fields. We matched the 846 stars to the fiducial spatial sample using the Gaia source identifier. Within the adopted kinematic core radius of $r\leq0.80~{\rm kpc}$ from each stellar-density peak, R1 contained 557 stars in the fiducial spatial sample, of which 106 had full six-dimensional information, giving a 6D sampling fraction of 19.0\%. R2 contained 368 stars, of which 118 had six-dimensional information, corresponding to 32.1\%. The $0.80~{\rm kpc}$ radius used for the kinematic analysis is independent of the $D_{80}$ characteristic spatial scale defined above.

We tested whether the positions of the 6D stars within each core differed systematically from those of the stars without full six-dimensional information. Radial distributions about the fixed stellar-density peak were compared using a two-sample Kolmogorov--Smirnov test, and the distributions of position angle were compared using a two-sample Kuiper statistic, with its null distribution estimated from 2,000 random permutations of the 6D and non-6D labels. For R1, the radial comparison gave a Kolmogorov--Smirnov statistic of 0.154 with $p=0.030$, while the position-angle comparison gave a Kuiper statistic of 0.193 with an empirical $p=0.029$. The R1 6D subset therefore showed modest spatial differences from the remainder of the fiducial sample. For R2, the corresponding values were 0.124 ($p=0.151$) for radius and 0.141 ($p=0.372$) for position angle, providing no comparable evidence for a spatial difference. These comparisons were used as diagnostics of spatial sampling non-uniformity rather than as formal hypothesis tests.

We therefore additionally tested the sensitivity of the inferred velocity-field geometry to the non-uniform spatial sampling of the 6D subset. Around each fixed R1 or R2 centre, we divided the fiducial spatial sample and the 6D subset into bins in radius and position angle, separately for the kinematic core and the local-background region defined below. For each bin containing 6D stars, we assigned those stars a weight proportional to
\begin{equation}
w_{ij}=\frac{N_{{\rm spatial},ij}}{N_{{\rm 6D},ij}},
\end{equation}
where $N_{{\rm spatial},ij}$ and $N_{{\rm 6D},ij}$ are the numbers of stars from the fiducial and 6D samples, respectively, in radial bin $i$ and position-angle bin $j$. The resulting stellar weights were normalised to have a mean of unity separately within each fitted region. We repeated the weighted analysis using $3\times6$, $4\times8$ and $5\times8$ radial-by-position-angle binning schemes. Because a bin containing fiducial stars but no 6D star cannot be represented by inverse weighting, we also recorded the fraction of the fiducial spatial sample lying in bins containing at least one 6D star.

For the nominal $4\times8$ scheme, we further performed 5,000 selection-weighted bootstrap realisations. Core and background stars were resampled separately with replacement, retaining the original number of stars in each region and assigning each 6D star a sampling probability proportional to its spatial-selection weight. The local background field and the core velocity-gradient tensor were then refitted in each realisation. The resulting sensitivity of the inferred velocity-field geometry to spatial variations in the 6D sampling is described below and shown in Extended Data Fig.~5.

\subsection{Local background velocity field}

The axisymmetric-rotation-corrected in-plane velocities defined above still contain spatially varying motions on scales larger than R1 and R2. To isolate the velocity field locally associated with each stellar overdensity, we estimated and subtracted an affine background velocity field fitted to the surrounding 6D stars. We adopted $R_{\rm core}=0.80~{\rm kpc}$ as a common kinematic aperture for R1 and R2. This aperture encloses 87.4\% and 87.5\% of the background-subtracted stellar excess in R1 and R2, respectively. Based on the background-subtracted radial profiles, we adopted outer radii of $1.25~{\rm kpc}$ for R1 and $1.00~{\rm kpc}$ for R2, and placed the inner edge of each kinematic-background annulus $0.10~{\rm kpc}$ beyond these radii, with a common radial width of $0.40~{\rm kpc}$. The resulting annuli are $1.35<r\leq1.75~{\rm kpc}$ for R1 and $1.10<r\leq1.50~{\rm kpc}$ for R2. These kinematic-background annuli are distinct from the spatial-background annuli used to determine $D_{80}$.

For each region, we defined
\begin{equation}
\Delta X=X-X_{\rm peak},\qquad
\Delta Y=Y-Y_{\rm peak},
\end{equation}
where $(X_{\rm peak},Y_{\rm peak})$ is the fixed R1 or R2 stellar-density peak. We fitted the two components of the local background velocity independently with the affine model
\begin{align}
u_{\rm bg}(\Delta X,\Delta Y)
&=
a_0+a_X\Delta X+a_Y\Delta Y,\\
v_{\rm bg}(\Delta X,\Delta Y)
&=
b_0+b_X\Delta X+b_Y\Delta Y.
\end{align}
The coefficients were determined by ordinary least squares using only the 6D stars in the corresponding background annulus. The constant terms give the velocity intercepts of the fitted background field at the adopted stellar-density peak, while the linear terms allow the surrounding velocity field to contain first-order spatial gradients that would not be removed by subtracting a single bulk velocity.

We evaluated the fitted background field at the position of every 6D star in the corresponding kinematic core and defined the background-subtracted velocities as
\begin{align}
u' &= u-u_{\rm bg}(\Delta X,\Delta Y),\\
v' &= v-v_{\rm bg}(\Delta X,\Delta Y).
\end{align}
The background model was determined entirely from stars outside the $0.80~{\rm kpc}$ core; no core-star velocities entered the fit. The use of an affine rather than constant background was intended to remove smooth local translation and first-order velocity gradients in the surrounding Galactic field before characterising the internal flow of each overdensity. We treated this affine field as a nuisance model for the smoothly varying local background and did not assign its fitted coefficients to a unique physical mechanism.

For quantities that require a spatial origin within the core, including radial velocities, we defined the stellar centroid from the positions of all stars in the fiducial spatial sample lying within $0.80~{\rm kpc}$ of the fixed density peak. No velocity information entered this centroid definition. The background-subtracted core velocities can retain a residual constant translation because the affine field is constrained by the surrounding annulus rather than by the core itself. For the displayed residual-velocity fields and the model-independent radial diagnostics, we therefore removed this remaining bulk translation by subtracting the mean values of $u'$ and $v'$ over the 6D core sample. In the radial-flow and velocity-gradient fits described below, the corresponding constant velocity components were instead included as free intercept terms in the models. Thus, any residual uniform translation of the core was absorbed by the fitted intercepts and therefore did not contribute to the inferred radial-flow amplitudes or velocity gradients.

\subsection{Radial-flow models and characteristic timescales}

We quantified the radial component of the background-subtracted velocity field using the stellar centroid defined above as the spatial origin. For each 6D core star, we defined
\begin{equation}
x=X-X_{\rm c},\qquad
y=Y-Y_{\rm c},\qquad
r=(x^2+y^2)^{1/2},
\end{equation}
and the corresponding radial unit vector
\begin{equation}
\boldsymbol{e}_r=\frac{1}{r}(x,y).
\end{equation}
Here $(X_{\rm c},Y_{\rm c})$ is the centroid of the fiducial spatial sample within $0.80~{\rm kpc}$ of the fixed stellar-density peak. The velocity vector entering the fits was the background-subtracted in-plane velocity $\boldsymbol{v}'=(u',v')$ defined above.

We fitted two radial-flow models for the primary comparison. The constant-speed radial model was
\begin{equation}
\boldsymbol{v}'=\boldsymbol{v}_0+V_{\rm out}\boldsymbol{e}_r,
\end{equation}
where $\boldsymbol{v}_0=(u_0,v_0)$ is a free residual bulk velocity and $V_{\rm out}$ is a radius-independent radial-flow amplitude. The homologous model was
\begin{equation}
\boldsymbol{v}'=\boldsymbol{v}_0+H(x,y),
\end{equation}
corresponding to a radial velocity $v_r=Hr$. Positive $V_{\rm out}$ or $H$ therefore denotes outward motion. We also fitted a bulk-only model and, as an additional diagnostic, a model containing both constant and homologous radial terms,
\begin{equation}
\boldsymbol{v}'=\boldsymbol{v}_0+\left(V_0+Hr\right)\boldsymbol{e}_r.
\end{equation}
All models were fitted by ordinary least squares to the two in-plane velocity components simultaneously. The free bulk terms ensure that any residual uniform translation remaining after subtraction of the local background field is separated from the radial-flow component.

We compared the constant-speed and homologous models using the Bayesian information criterion (BIC) \cite{Schwarz1978}. For $N$ stars, the two velocity components give $n=2N$ fitted data values. We calculated
\begin{equation}
{\rm BIC}
=
n\ln\left(\frac{{\rm RSS}}{n}\right)
+k\ln n,
\end{equation}
where ${\rm RSS}$ is the summed squared residual over both velocity components and $k$ is the number of fitted parameters. We defined
\begin{equation}
\Delta{\rm BIC}
\equiv
{\rm BIC}_{\rm const}-{\rm BIC}_{\rm hom},
\end{equation}
so that $\Delta{\rm BIC}>0$ favours the homologous model and $\Delta{\rm BIC}<0$ favours the constant-speed model. The nominal values are $\Delta{\rm BIC}=-6.85$ for R1 and $+5.44$ for R2. As described below, measurement-error propagation and bootstrap resampling broaden these comparisons sufficiently that the 95\% joint intervals include zero for both regions; we therefore do not regard either radial-model preference as robust.

After removing the residual bulk velocity of the 6D core sample, we calculated radial diagnostics that do not assume either of the fitted radial-flow models. Defining
\begin{equation}
\tilde{u}=u'-\overline{u'},\qquad
\tilde{v}=v'-\overline{v'},
\end{equation}
the radial velocity of each star was
\begin{equation}
v_r=\frac{\tilde{u}x+\tilde{v}y}{r}.
\end{equation}
We used the median of $v_r$ to characterise the typical radial motion and the fraction of stars with $v_r>0$ as a non-parametric measure of the prevalence of outward motion. These diagnostics give outward fractions of 74.5\% and 86.4\%, and median radial velocities of $6.38$ and $7.53~{\rm km\,s^{-1}}$, for R1 and R2, respectively.

We used the fitted radial-flow amplitudes to define two characteristic kinematic timescales. For the constant-speed model, we defined
\begin{equation}
t_{r/V}\equiv\frac{D_{80}/2}{V_{\rm out}},
\end{equation}
using the fiducial characteristic radius $D_{80}/2$. For the homologous model, the corresponding characteristic timescale is
\begin{equation}
t_H\equiv H^{-1}.
\end{equation}
The conversion $1~{\rm kpc}/(1~{\rm km\,s^{-1}})=977.8~{\rm Myr}$ was used. These quantities are dimensional timescales inferred from the present-day velocity field and are not interpreted as dynamical ages or as evidence for a single coeval expansion event.

The quoted timescale ranges were obtained by inverse-transforming the median and central 95\% intervals of the joint measurement-error and bootstrap distributions of $V_{\rm out}$ and $H$ described below, while holding the fiducial $D_{80}$ values fixed. The central timescale was calculated from the median flow amplitude. Because both timescales vary inversely with the corresponding amplitude, the ordering of the interval limits was reversed: the lower timescale limit was calculated from the 97.5th-percentile amplitude and the upper limit from the 2.5th-percentile amplitude. The 2.5th-percentile values of both $V_{\rm out}$ and $H$ are positive for R1 and R2, so these inverse-transformed amplitude intervals yield finite characteristic-timescale ranges. They should not be interpreted as percentile intervals of a separately constructed timescale distribution. This gives $t_{r/V}=92.7^{+82.1}_{-29.0}~{\rm Myr}$ and $t_H=73.2^{+105.3}_{-26.8}~{\rm Myr}$ for R1, and $t_{r/V}=78.0^{+54.2}_{-24.8}~{\rm Myr}$ and $t_H=43.5^{+40.5}_{-16.3}~{\rm Myr}$ for R2.

\subsection{Two-dimensional velocity-gradient tensor}

To characterise the geometry of the internal velocity field without imposing radial symmetry, we fitted a general first-order velocity field to the background-subtracted motions of the 6D core stars. Using the same centroid-based coordinates $(x,y)$ defined above, we modelled
\begin{equation}
\begin{pmatrix}
u'\\
v'
\end{pmatrix}
=
\begin{pmatrix}
u_0\\
v_0
\end{pmatrix}
+
\mathbf{A}
\begin{pmatrix}
x\\
y
\end{pmatrix},
\end{equation}
where $(u_0,v_0)$ is a free residual bulk velocity and
\begin{equation}
\mathbf{A}
=
\begin{pmatrix}
\partial u'/\partial x & \partial u'/\partial y\\
\partial v'/\partial x & \partial v'/\partial y
\end{pmatrix}
\end{equation}
is the two-dimensional velocity-gradient tensor. The six parameters comprising the two intercepts and four velocity gradients were fitted by ordinary least squares to the two in-plane velocity components simultaneously. No radial, axisymmetric or isotropic form was imposed on the internal velocity field; this first-order velocity-gradient representation is analogous to linear stellar-kinematic formalisms used in Galactic velocity-field studies \cite{Fedorov2021}.

We decomposed the velocity-gradient tensor into symmetric and antisymmetric components,
\begin{equation}
\mathbf{A}
=
\mathbf{S}+\mathbf{\Omega},
\qquad
\mathbf{S}
=
\frac{1}{2}\left(\mathbf{A}+\mathbf{A}^{\rm T}\right),
\qquad
\mathbf{\Omega}
=
\frac{1}{2}\left(\mathbf{A}-\mathbf{A}^{\rm T}\right).
\end{equation}
The divergence of the fitted field was
\begin{equation}
\nabla\cdot\boldsymbol{v}'
=
{\rm tr}(\mathbf{A})
=
\frac{\partial u'}{\partial x}
+
\frac{\partial v'}{\partial y},
\end{equation}
and the out-of-plane vorticity was defined as
\begin{equation}
\omega_z
=
\frac{\partial v'}{\partial x}
-
\frac{\partial u'}{\partial y}.
\end{equation}
Both quantities have units of ${\rm km\,s^{-1}\,kpc^{-1}}$. Positive divergence corresponds to net local expansion in the fitted plane, whereas $\omega_z$ describes its local rotational component.

We quantified directional expansion using the eigenvalues of the symmetric rate-of-strain tensor $\mathbf{S}$. We ordered these as
\begin{equation}
\lambda_1\geq\lambda_2,
\end{equation}
with the associated eigenvectors defining the principal strain axes. A positive eigenvalue denotes extension along its corresponding principal axis and a negative eigenvalue denotes contraction. Because
\begin{equation}
\lambda_1+\lambda_2
=
{\rm tr}(\mathbf{S})
=
\nabla\cdot\boldsymbol{v}',
\end{equation}
the two principal strain rates also specify how the total divergence is distributed between the two principal directions. For reference, an isotropic homologous expansion would give $\lambda_1=\lambda_2>0$, whereas a strongly anisotropic divergent field can have one positive principal strain rate and a second value close to zero.

For the nominal data, the fitted principal strain rates are $\lambda_1=25.4$ and $\lambda_2=1.6~{\rm km\,s^{-1}\,kpc^{-1}}$ for R1, and $\lambda_1=28.2$ and $\lambda_2=24.0~{\rm km\,s^{-1}\,kpc^{-1}}$ for R2. The corresponding nominal divergences are $27.0$ and $52.2~{\rm km\,s^{-1}\,kpc^{-1}}$, respectively. These values refer to the fixed $0.80~{\rm kpc}$ core and the region-specific local-background subtraction defined above. Their uncertainties, including measurement-error propagation and bootstrap resampling, and the sensitivity of the inferred geometry to the adopted kinematic selections are quantified below.

\subsection{Measurement uncertainties and bootstrap resampling}

To propagate measurement uncertainties and finite-sample variation through the kinematic analysis, we generated 5,000 joint measurement-error and bootstrap realisations. Distance uncertainties were propagated for every star in the fiducial spatial sample using the median, lower and upper photogeometric distance estimates described above. For each star, we defined
\begin{equation}
\sigma_-=\ln\left(\frac{d_{\rm med}}{d_{\rm lo}}\right),
\qquad
\sigma_+=\ln\left(\frac{d_{\rm hi}}{d_{\rm med}}\right),
\end{equation}
and drew a standard-normal variate $q$. The perturbed distance was then
\begin{equation}
d=
\begin{cases}
d_{\rm med}\exp(q\sigma_-), & q<0,\\
d_{\rm med}\exp(q\sigma_+), & q\geq0.
\end{cases}
\end{equation}
This asymmetric log-distance prescription reproduces the reported lower and upper distance estimates at $q=-1$ and $q=+1$, respectively. The corresponding distance draw for each 6D star was inherited from its source-identifier match to the fiducial spatial sample.

For the 6D subset, the two proper-motion components and the radial velocity were independently perturbed with zero-mean Gaussian deviates using their quoted measurement uncertainties. Sky positions were held fixed. For every realisation, the perturbed distances, proper motions and radial velocities were used to recalculate the Galactocentric positions and space velocities, followed by the same solar-motion and axisymmetric-rotation corrections adopted for the nominal analysis.

The R1 and R2 stellar-density peak coordinates, the core radius of $0.80~{\rm kpc}$ and the region-specific kinematic-background annuli were held fixed. Because the stellar positions changed after perturbing the distances, membership of the 6D core and background samples was reassigned in every realisation. The spatial centroid used as the origin of the internal velocity field was likewise recalculated from all stars in the perturbed fiducial spatial sample lying within $0.80~{\rm kpc}$ of the corresponding fixed density peak. A realisation was retained only if it contained at least 20 6D core stars, 10 6D background stars and 20 fiducial spatial-sample stars in the core; all 5,000 realisations satisfied these criteria for both R1 and R2.

After the measurement perturbations and sample reassignment, the 6D core and background samples were resampled separately with replacement, preserving the number of stars in each sample for that realisation. The local affine background velocity field was refitted to the resampled background stars and evaluated at the positions of the resampled core stars. We then repeated the complete internal-kinematic analysis on the resulting background-subtracted core velocities. In each realisation we refitted the constant-speed and homologous radial-flow models, recalculated the median radial velocity and outward-moving fraction, and evaluated
\begin{equation}
\Delta{\rm BIC}
=
{\rm BIC}_{\rm const}-{\rm BIC}_{\rm hom}.
\end{equation}
We also refitted the full two-dimensional velocity-gradient tensor and recalculated its divergence, vorticity and principal strain rates $\lambda_1$ and $\lambda_2$. Thus, uncertainty in stellar distances affects not only the inferred velocities but also core and background membership, the spatial centroid and the fitted local background field.

For each derived quantity, we summarised the 5,000 joint realisations by the median and the 16th--84th and 2.5th--97.5th percentile ranges, which we refer to as the central 68\% and 95\% joint uncertainty intervals. These intervals incorporate both the adopted measurement-error perturbations and finite-sample variation represented by bootstrap resampling. For quantities with a natural reference value, we also calculated the fraction of realisations exceeding that value: zero for signed quantities, including $\Delta{\rm BIC}$, and 0.5 for the outward-moving fraction.

For the radial-flow fits, the joint distributions give $V_{\rm out}=6.7~{\rm km\,s^{-1}}$ with a 95\% interval of $3.6$--$9.7~{\rm km\,s^{-1}}$ and $H=13.4~{\rm km\,s^{-1}\,kpc^{-1}}$ with a 95\% interval of $5.5$--$21.1~{\rm km\,s^{-1}\,kpc^{-1}}$ for R1. The corresponding values for R2 are $V_{\rm out}=8.1~{\rm km\,s^{-1}}$ with a 95\% interval of $4.8$--$12.0~{\rm km\,s^{-1}}$ and $H=22.5~{\rm km\,s^{-1}\,kpc^{-1}}$ with a 95\% interval of $11.6$--$36.0~{\rm km\,s^{-1}\,kpc^{-1}}$. The median $\Delta{\rm BIC}$ values are $-5.37$ for R1 and $+4.73$ for R2, but their 95\% intervals, $-17.19$ to $+6.17$ and $-6.70$ to $+21.72$, respectively, both include zero.

For the velocity-gradient tensor, R1 gives a joint median divergence of $29.0~{\rm km\,s^{-1}\,kpc^{-1}}$ with a 95\% interval of $9.1$--$45.9~{\rm km\,s^{-1}\,kpc^{-1}}$. Its principal strain rates are $\lambda_1=27.2~{\rm km\,s^{-1}\,kpc^{-1}}$ with a 95\% interval of $14.4$--$39.3~{\rm km\,s^{-1}\,kpc^{-1}}$ and $\lambda_2=1.7~{\rm km\,s^{-1}\,kpc^{-1}}$ with a 95\% interval of $-8.9$--$12.0~{\rm km\,s^{-1}\,kpc^{-1}}$. For R2, the corresponding divergence is $47.3~{\rm km\,s^{-1}\,kpc^{-1}}$ with a 95\% interval of $22.8$--$77.8~{\rm km\,s^{-1}\,kpc^{-1}}$. The principal strain rates are $\lambda_1=31.0~{\rm km\,s^{-1}\,kpc^{-1}}$ and $\lambda_2=15.7~{\rm km\,s^{-1}\,kpc^{-1}}$, with 95\% intervals of $16.1$--$57.3$ and $2.3$--$28.6~{\rm km\,s^{-1}\,kpc^{-1}}$, respectively. These joint distributions provide the uncertainty intervals used for the kinematic quantities reported in the main text and Fig.~4.

\subsection{Kinematic robustness tests}

We tested the sensitivity of the inferred velocity-field geometry to the adopted core aperture and local-background region. The core radius was varied over $R_{\rm core}=0.70$, $0.80$ and $0.90~{\rm kpc}$. For each core radius, the fiducial kinematic-background annulus was shifted inward or outward by $0.10~{\rm kpc}$ while retaining its width of $0.40~{\rm kpc}$. The three background ranges were therefore $1.25<r\leq1.65$, $1.35<r\leq1.75$ and $1.45<r\leq1.85~{\rm kpc}$ for R1, and $1.00<r\leq1.40$, $1.10<r\leq1.50$ and $1.20<r\leq1.60~{\rm kpc}$ for R2. For every combination, we refitted the local affine background field and the two-dimensional velocity-gradient tensor. Across the nine combinations for each region, R1 retained a strongly unequal pair of principal strain rates, with $\lambda_1=22.3$--$34.5$ and $\lambda_2=0.7$--$6.7~{\rm km\,s^{-1}\,kpc^{-1}}$, whereas R2 retained two positive principal strain rates, with $\lambda_1=20.7$--$40.4$ and $\lambda_2=13.8$--$24.5~{\rm km\,s^{-1}\,kpc^{-1}}$. The fitted divergence remained positive for all tested settings, spanning $23.5$--$40.9~{\rm km\,s^{-1}\,kpc^{-1}}$ for R1 and $34.4$--$64.4~{\rm km\,s^{-1}\,kpc^{-1}}$ for R2.

We also tested sensitivity to individual stars and to stars with large residual velocity amplitudes. In a leave-one-out analysis, each core star and each background star was removed in turn and the corresponding analysis was repeated, refitting the local background field when a background star was removed and refitting the internal velocity-gradient tensor in all cases. For R1, removing individual core stars gave $\lambda_2=-1.3$ to $2.9~{\rm km\,s^{-1}\,kpc^{-1}}$, while removing individual background stars gave $0.9$ to $2.5~{\rm km\,s^{-1}\,kpc^{-1}}$. The corresponding ranges for R2 were $16.6$--$26.0$ and $23.4$--$24.8~{\rm km\,s^{-1}\,kpc^{-1}}$, respectively. We further removed the stars with the largest residual velocity amplitudes separately from the core and background samples, using trimming fractions of 1, 2.5, 5 and 10\%. Even after removing 10\% of the stars in each sample, the same qualitative pattern remained: over the tested trimming fractions, $\lambda_2$ remained between $0.5$ and $1.5~{\rm km\,s^{-1}\,kpc^{-1}}$ for R1 and between $15.5$ and $18.1~{\rm km\,s^{-1}\,kpc^{-1}}$ for R2.

Because the 6D subset is not spatially uniform within the fiducial massive-star sample, we additionally repeated the tensor analysis using the spatial-selection weights defined above. We used $3\times6$, $4\times8$ and $5\times8$ radial-by-position-angle binning schemes. Across these three schemes, the weighted fits gave $\lambda_2=-1.8$ to $1.2~{\rm km\,s^{-1}\,kpc^{-1}}$ for R1 and $12.9$--$23.0~{\rm km\,s^{-1}\,kpc^{-1}}$ for R2. For the nominal $4\times8$ scheme, bins containing at least one 6D star represented 98.0\% and 92.9\% of the fiducial core samples of R1 and R2, respectively; the corresponding coverage fractions in the background regions were 81.0\% and 97.4\%. Because fiducial-sample stars in bins containing no 6D star cannot be recovered by inverse weighting, we regard this procedure as a sensitivity test rather than a complete correction for 6D selection effects.

We then performed 5,000 selection-weighted bootstrap realisations using the $4\times8$ scheme. These selection-weighted bootstrap realisations were performed as a separate sampling-sensitivity test rather than as part of the joint measurement-error propagation described above. Core and background stars were resampled separately with replacement, with sampling probabilities proportional to their spatial-selection weights, and the local background field and velocity-gradient tensor were refitted in each realisation. The resulting median principal strain rates were $\lambda_1=24.3$ and $\lambda_2=0.6~{\rm km\,s^{-1}\,kpc^{-1}}$ for R1, with central 95\% bootstrap intervals of $12.8$--$38.5$ and $-10.8$--$12.2~{\rm km\,s^{-1}\,kpc^{-1}}$, respectively. For R2, the corresponding values were $\lambda_1=25.3$ and $\lambda_2=15.1~{\rm km\,s^{-1}\,kpc^{-1}}$, with central 95\% bootstrap intervals of $15.5$--$42.1$ and $5.2$--$25.8~{\rm km\,s^{-1}\,kpc^{-1}}$, respectively. In 97.0\% of these weighted bootstrap realisations, $\lambda_2$ was larger in R2 than in R1. The central 95\% bootstrap interval of the direct difference nevertheless extended slightly below zero, so this fraction is used as a descriptive measure of robustness rather than as a formal significance statement.

Finally, we tested whether the R1 radial-flow signal was driven by stars matched to the published Car~OB1 sample \cite{Goppl2025}. Nine stars in the fiducial massive-star sample were matched by Gaia DR3 source identifier to the published Car~OB1 sample, of which only one lay in the R1 6D core. Removing all nine matched stars changed the number of R1 core stars with 6D information from 106 to 105. With the original spatial centroid retained, the fitted constant radial-flow amplitude changed from $6.59$ to $6.57~{\rm km\,s^{-1}}$, the median radial velocity from $6.38$ to $6.37~{\rm km\,s^{-1}}$, and the outward-moving fraction from 74.5\% to 75.2\%. Recalculating the spatial centroid after the exclusion produced similarly small changes. The large-scale R1 radial-flow signal is therefore not driven by the matched published Car~OB1 members.

Together, these tests show that the qualitative pattern of a strongly anisotropic divergent field in R1 and a two-dimensional divergent field in R2 is not driven by the adopted kinematic apertures or background regions, individual stars, stars with the largest residual velocities, or spatial variations in the sampling of the 6D subset. The exclusion of stars matched to the published Car~OB1 sample further shows that the large-scale R1 radial-flow signal is not driven by these matched members. The robustness of the inferred velocity-field geometries to the aperture and background choices, leave-one-out and high-residual-velocity trimming tests, and spatial-selection weighting is summarised in Extended Data Fig.~5.

\subsection{Use of large language models}

OpenAI ChatGPT was used interactively by the first author during the preparation of this study to assist with drafting and editing Python analysis code, debugging and consistency checks, consideration of alternative methodological approaches and robustness tests, and drafting and language editing of the manuscript. Prompts included requests to inspect or revise author-provided analysis code and text, to suggest candidate robustness tests, and to improve the clarity and internal consistency of the manuscript. AI-generated suggestions, code and text incorporated into the study were reviewed and, where necessary, modified by the first author. All analyses reported in this study were executed by the first author, numerical results were verified against the executed analysis code, and the final methodological choices, scientific interpretations and conclusions were determined by the authors. No generative-AI image-generation system was used to create or alter the scientific figures in this manuscript.

\subsection{Data availability}

The input data products used in this study are publicly available from the Gaia Data Release 3 archive and from the published catalogues and three-dimensional extinction data cited in the Methods. The derived massive-star catalogue used in the analysis, including the fiducial 3,502-star spatial sample and the 846-star six-dimensional kinematic subset, together with the numerical source data underlying the main and Extended Data figures, will be deposited in a public repository no later than publication. These derived data will be made available to editors and referees upon request during peer review.

\subsection{Code availability}

Custom Python code used to construct the massive-star sample, identify and characterise R1 and R2, analyse their association with the three-dimensional extinction field, calculate stellar kinematics, fit the radial-flow and velocity-gradient models, propagate measurement uncertainties, perform the robustness tests, and generate the figures will be deposited in a public repository no later than publication. The code will be made available to editors and referees upon request during peer review.

\backmatter

\bmhead{Acknowledgements}

This work has made use of data from the European Space Agency (ESA) mission {\it Gaia} (\url{https://www.cosmos.esa.int/gaia}), processed by the {\it Gaia} Data Processing and Analysis Consortium (DPAC, \url{https://www.cosmos.esa.int/web/gaia/dpac/consortium}). Funding for the DPAC has been provided by national institutions, in particular the institutions participating in the {\it Gaia} Multilateral Agreement.

\bmhead{Funding}

This work received no specific funding.

\bmhead{Author contributions}

S.N. developed the study, performed the data analysis, developed and implemented the analysis methods and code, produced the figures, interpreted the results, and wrote the manuscript. K.T. contributed to the conception of the study by proposing the application of the previously developed massive-star selection method to the Milky Way using a three-dimensional dust map, and contributed to the scientific interpretation of the results and substantive revision of the manuscript. Both authors approved the submitted version of the manuscript.

\bmhead{Competing interests}

The authors declare no competing interests.

\bmhead{Additional information}

Supplementary information is available for this paper.

Correspondence and requests for materials should be addressed to S.N.

\clearpage

\setcounter{figure}{0}
\renewcommand{\figurename}{Extended Data Fig.}
\makeatletter
\renewcommand{\theHfigure}{ED.\arabic{figure}}
\makeatother

\begin{figure}[p]
\centering
\includegraphics[width=\textwidth]{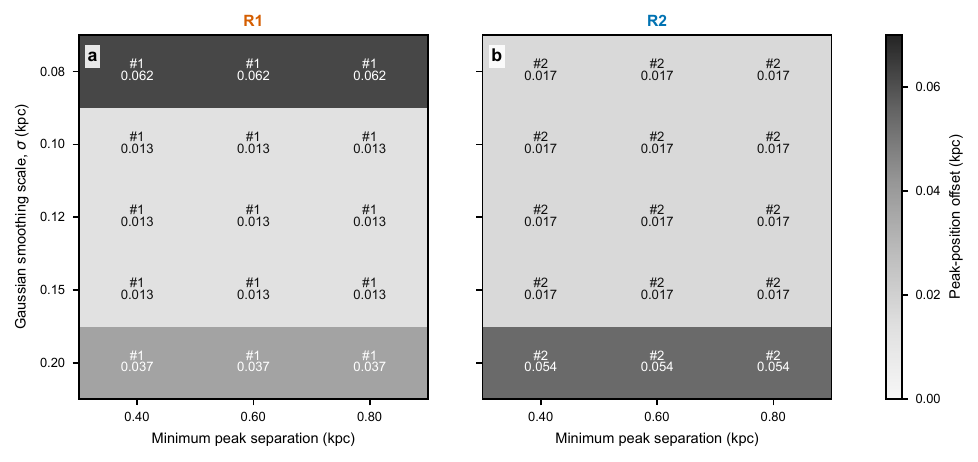}
\caption{\textbf{Robustness of the density-based identification of R1 and R2.}
\textbf{a,b,} Peak ranks and positional offsets for R1 (\textbf{a}) and R2 (\textbf{b}) obtained by varying the Gaussian smoothing scale, $\sigma$, and the minimum separation between distinct local maxima in the stellar surface-density field. The search was repeated for five smoothing scales, $\sigma=0.08$, 0.10, 0.12, 0.15 and $0.20~{\rm kpc}$, and three minimum peak separations, 0.40, 0.60 and $0.80~{\rm kpc}$, giving 15 parameter combinations. For each combination, the highest-ranked detected peak within $0.50~{\rm kpc}$ of the adopted R1 or R2 centre was matched to that structure, solely to track the same structures across the parameter grid. In each cell, the upper value gives the rank of the matched peak by smoothed stellar density among the distinct local maxima, and the lower value gives its positional offset from the adopted centre in kpc. Greyscale shading encodes the same positional offset, with darker cells indicating larger shifts. R1 remains the highest-ranked peak and R2 the second-highest-ranked peak in all 15 combinations. The maximum positional offsets are $0.062~{\rm kpc}$ for R1 and $0.054~{\rm kpc}$ for R2, so all recovered peak positions for both structures remain within $0.07~{\rm kpc}$ of their adopted centres across the full parameter grid.}
\label{fig:ED1}
\end{figure}

\clearpage
\begin{figure}[p]
\centering
\includegraphics[width=\textwidth]{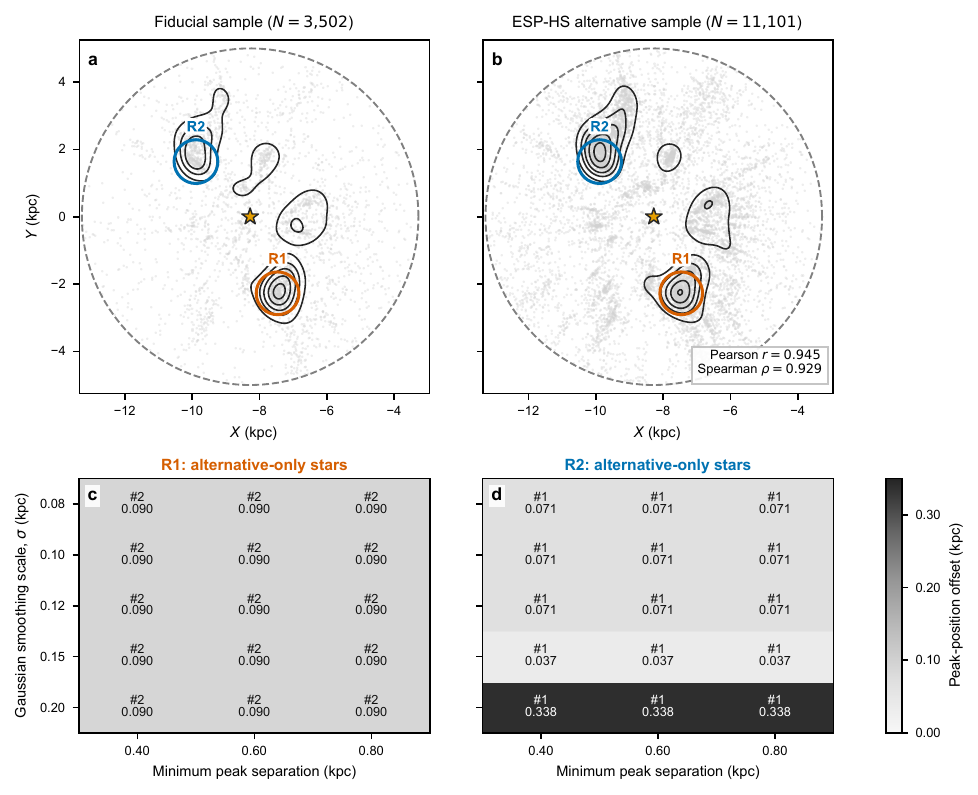}
\caption{\textbf{Robustness of R1 and R2 to an alternative extinction treatment.}
\textbf{a,b,} Smoothed stellar surface-density fields for the fiducial sample of 3,502 massive-star candidates (\textbf{a}) and the alternative sample of 11,101 candidates obtained using Gaia DR3 ESP-HS extinction estimates where available (\textbf{b}). Both fields were evaluated on the same $0.05~{\rm kpc}$ grid and smoothed with $\sigma=0.30~{\rm kpc}$. Contours show normalised density levels of 0.20, 0.35, 0.50, 0.70 and 0.90. Orange and blue circles mark the adopted R1 and R2 $D_{80}$ diameters, respectively, and the yellow star marks the Sun. Within $5~{\rm kpc}$ of the Sun, for grid cells in which at least one field is non-zero, the smoothed fields yield Pearson $r=0.945$ and Spearman $\rho=0.929$.
\textbf{c,d,} Peak robustness for R1 (\textbf{c}) and R2 (\textbf{d}) using only the 8,004 ESP-HS-selected candidates absent from the fiducial sample. The density search was repeated over the same five smoothing scales and three minimum peak separations as in Extended Data Fig.~\ref{fig:ED1}, giving 15 combinations. Peaks were matched to the fiducial centres using the same $0.50~{\rm kpc}$ tracking criterion. In each cell, the upper value gives the peak rank by smoothed stellar density and the lower value its positional offset from the fiducial centre; greyscale shading encodes the same offset. R1 remains the second-highest-ranked peak in all 15 combinations and remains within $0.090~{\rm kpc}$ of its fiducial centre. R2 remains the highest-ranked peak throughout; its offset is at most $0.071~{\rm kpc}$ for $\sigma\leq0.12~{\rm kpc}$, $0.037~{\rm kpc}$ at $\sigma=0.15~{\rm kpc}$ and $0.338~{\rm kpc}$ at $\sigma=0.20~{\rm kpc}$. Thus, both overdensities persist under the alternative extinction treatment, although the R2 peak position becomes sensitive to the strongest smoothing.}
\label{fig:ED2}
\end{figure}

\clearpage

\begin{figure}[p]
\centering
\includegraphics[width=\textwidth]{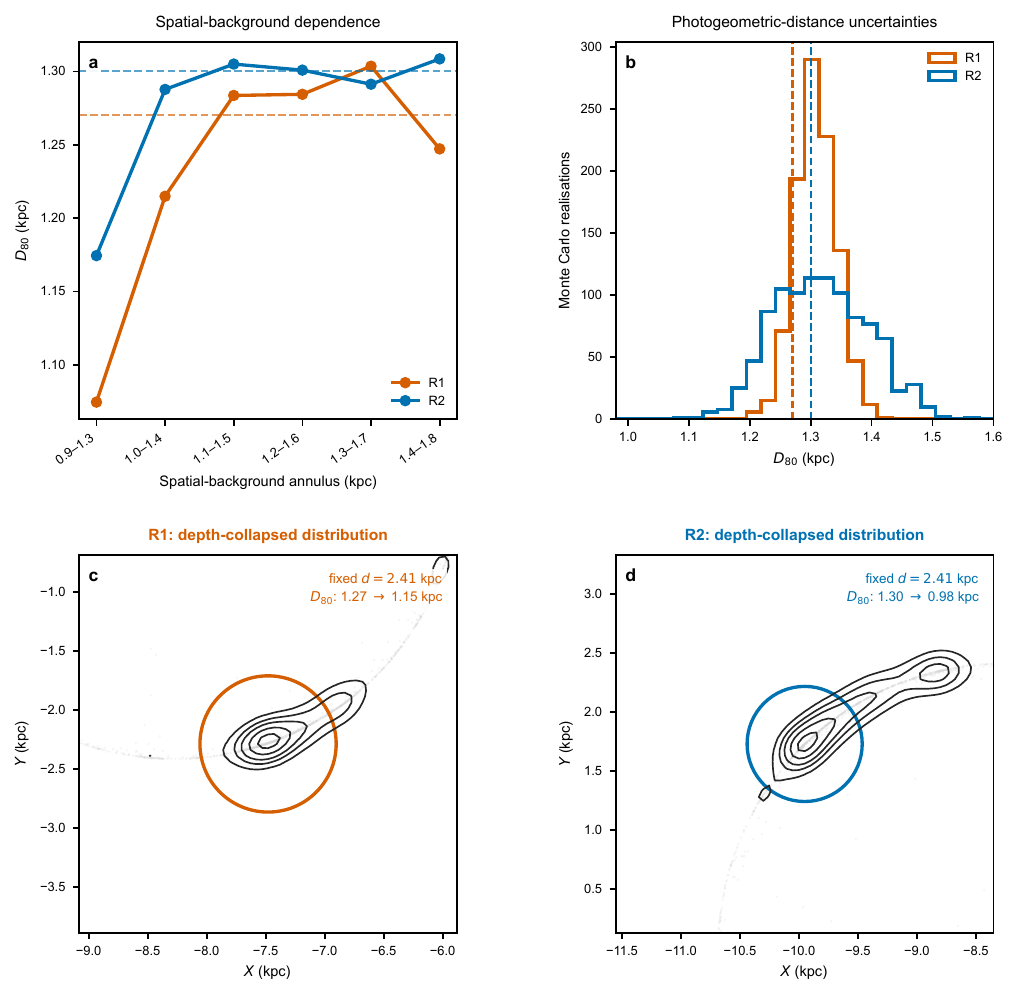}
\caption{\textbf{Robustness of the characteristic spatial scales of R1 and R2.}
\textbf{a,} Dependence of $D_{80}$ on the spatial-background annulus used to estimate the local background level of the stellar surface density. Orange and blue points show R1 and R2, respectively; dashed lines mark the adopted values, $D_{80}=1.27~{\rm kpc}$ for R1 and $1.30~{\rm kpc}$ for R2. Across the six background choices, $D_{80}$ ranges from 1.08 to $1.30~{\rm kpc}$ for R1 and from 1.17 to $1.31~{\rm kpc}$ for R2.
\textbf{b,} Propagation of photogeometric-distance uncertainties into $D_{80}$ from 1,000 Monte Carlo realisations. The median values are 1.31 and $1.32~{\rm kpc}$ for R1 and R2, respectively, with central 95\% intervals of 1.25--$1.37~{\rm kpc}$ and 1.19--$1.47~{\rm kpc}$. Dashed lines mark the adopted fiducial values.
\textbf{c,d,} Limiting tests in which the line-of-sight depth is removed by placing the full spatial sample at a single region-specific heliocentric distance while preserving each star's sky position, separately for R1 (\textbf{c}) and R2 (\textbf{d}). The distances were determined independently for the two regions and are both approximately $2.41~{\rm kpc}$; no common distance was imposed. Grey points show the resulting depth-collapsed stellar positions and black contours the smoothed stellar-density field. The orange and blue circles are centred on the reidentified density peaks and have diameters equal to the resulting $D_{80}$ values. The inferred sizes decrease from 1.27 to $1.15~{\rm kpc}$ for R1 and from 1.30 to $0.98~{\rm kpc}$ for R2, showing that both structures remain extended on scales of approximately $1~{\rm kpc}$ after removal of the line-of-sight depth.}
\label{fig:ED3}
\end{figure}

\clearpage

\begin{figure}[p]
\centering
\includegraphics[width=\textwidth]{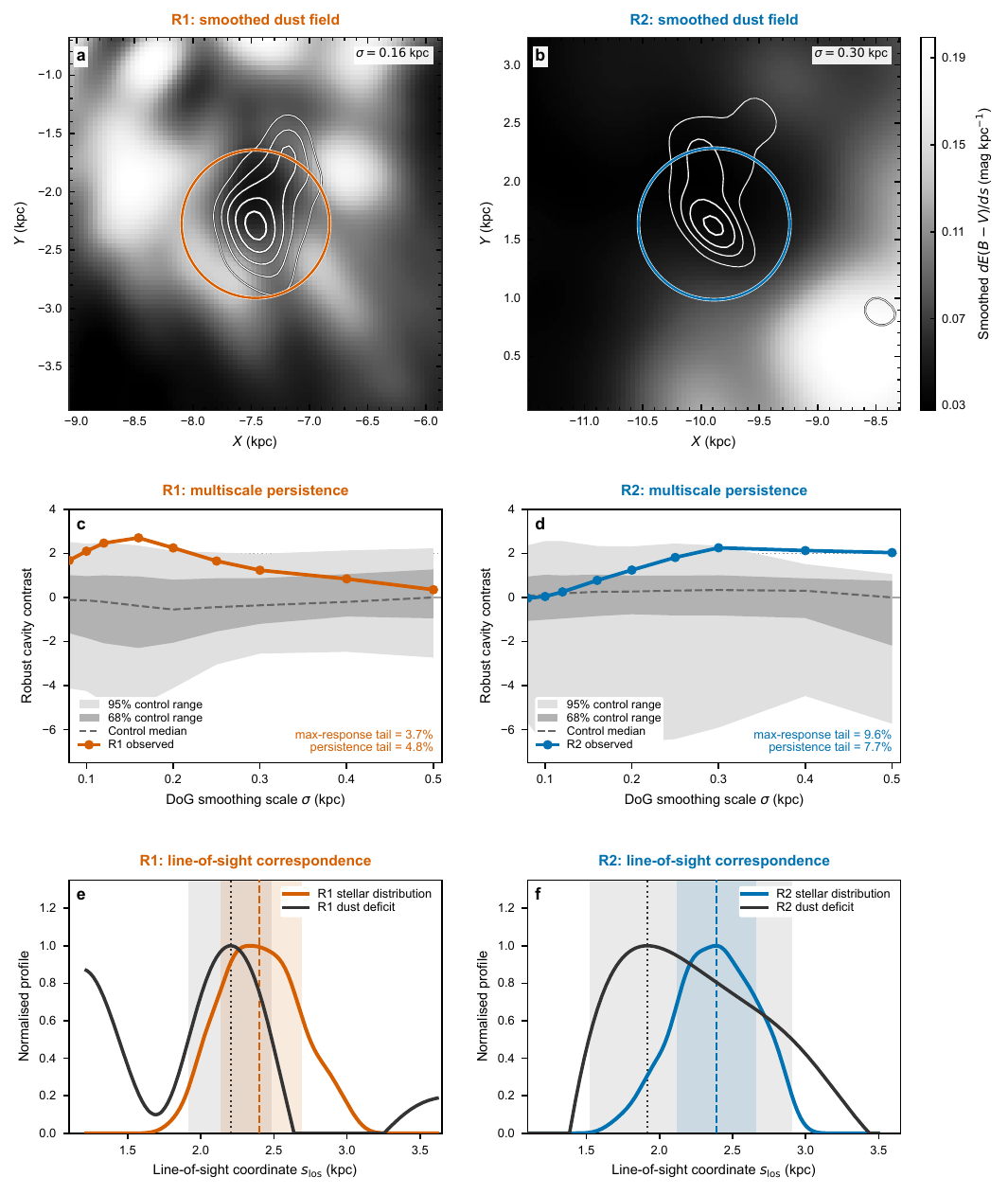}
\caption{\textbf{Three-dimensional stellar--dust correspondence for R1 and R2.}
\textbf{a,b,} Smoothed differential reddening, $dE(B-V)/ds$, around R1 (\textbf{a}) and R2 (\textbf{b}) from the three-dimensional dust-extinction map \cite{Wang2025Dust}, averaged over $|Z|\leq0.30~{\rm kpc}$ and Gaussian-smoothed with $\sigma=0.16$ and $0.30~{\rm kpc}$, respectively. Darker regions indicate lower differential reddening. White contours show the normalised massive-star surface-density field at levels of 0.20, 0.35, 0.50, 0.70 and 0.90, as in Fig.~\ref{fig:discovery}. The orange and blue circles are centred on the R1 and R2 stellar-density peaks and have diameters $D_{80}=1.27$ and $1.30~{\rm kpc}$, respectively. Both panels use the same greyscale range; a monotonic contrast enhancement was applied for display, while the colour bar shows the underlying smoothed $dE(B-V)/ds$ values.
\textbf{c,d,} Multiscale dust-deficit responses at the fixed R1 (\textbf{c}) and R2 (\textbf{d}) stellar-density peaks. Coloured curves show the observed robust cavity contrast from the difference-of-Gaussians (DoG) analysis, dashed grey curves the medians of local control positions, and dark- and light-grey bands their central 68\% and 95\% ranges. Positive contrast indicates a local dust deficit; the dotted line marks a contrast of 2. The max-response tail is the fraction of control positions with an equal or larger maximum response across scales, and the persistence tail is the fraction with an equal or longer consecutive run above 2. These are empirical spatial tail fractions, not formal $P$ values.
\textbf{e,f,} Line-of-sight correspondence for R1 (\textbf{e}) and R2 (\textbf{f}) at the smoothing scales producing the strongest projected dust-deficit responses. Coloured curves show the normalised stellar $s_{\rm los}$ distributions and dark curves the normalised dust-deficit profiles. Coloured shading marks the central 68\% interval of the stellar distribution and grey shading the half-maximum interval of the dust-deficit profile. Coloured dashed lines mark the stellar medians and dark dotted lines the dust-deficit peaks.}
\label{fig:ED4}
\end{figure}

\clearpage

\begin{figure}[p]
\centering
\includegraphics[width=\textwidth]{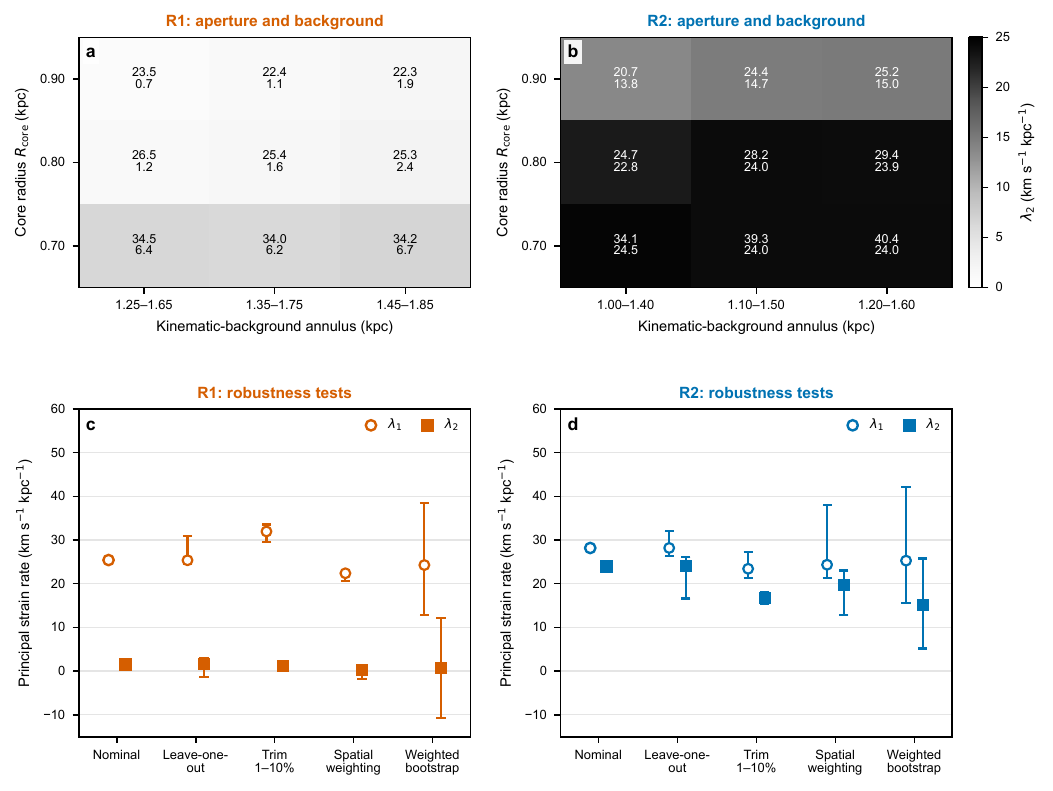}
\caption{\textbf{Robustness of the contrasting velocity-field geometries in R1 and R2.}
\textbf{a,b,} Principal strain rates for R1 (\textbf{a}) and R2 (\textbf{b}) across the nine tested combinations of kinematic core radius and local-background annulus. In each cell, the upper and lower values give $\lambda_1$ and $\lambda_2$, respectively, in ${\rm km\,s^{-1}\,kpc^{-1}}$; greyscale shading encodes $\lambda_2$.
\textbf{c,d,} Principal strain rates for R1 (\textbf{c}) and R2 (\textbf{d}) for the nominal analysis and the leave-one-out, residual-velocity trimming, spatial-selection weighting and selection-weighted bootstrap tests. Open circles and filled squares denote $\lambda_1$ and $\lambda_2$, respectively, and the horizontal grey line marks zero strain. For the leave-one-out, 1--10\% residual-velocity trimming and spatial-selection weighting tests, vertical bars span the full range across the tested variants. For the selection-weighted bootstrap, points show the medians and vertical bars the central 95\% intervals of 5,000 realisations. R1 retains a strongly unequal pair of principal strain rates across the tests, whereas R2 retains two positive principal strain rates.}
\label{fig:ED5}
\end{figure}

\clearpage

\bibliography{sn-bibliography}

\end{document}


\begin{center}
{\Large\bfseries Supplementary Information}

\vspace{1em}

{\large Contrasting geometries of kiloparsec-scale stellar flows associated with giant Galactic cavities}

\vspace{0.5em}

Satoya Nakano and Kengo Tachihara
\end{center}

\clearpage

\begin{figure}[p]
\centering
\includegraphics[width=\textwidth]{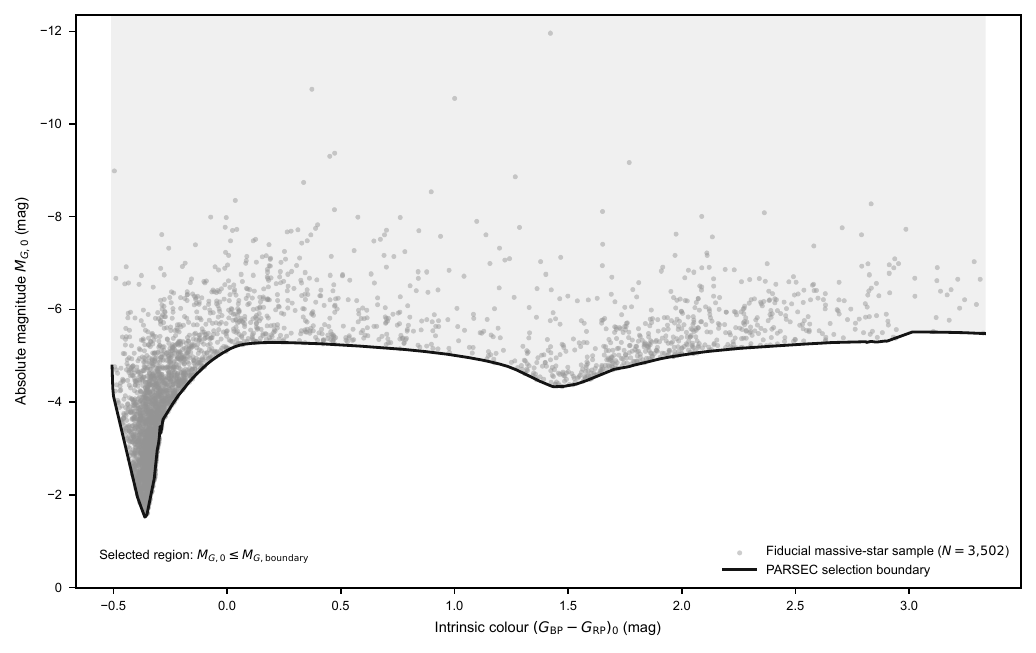}
\caption{\textbf{Extinction-corrected colour--magnitude diagram and massive-star selection.}
Grey points show the 3,502 candidates in the fiducial massive-star sample in intrinsic Gaia colour and extinction-corrected absolute magnitude. The black curve shows the selection boundary constructed from PARSEC v1.2S models at $Z_{\rm ini}=0.0152$ to encompass the evolutionary loci of stars with initial masses $M_{\rm ini}\gtrsim8~M_{\odot}$. The light-grey region denotes the adopted selection region, $M_{G,0}\leq M_{G,\rm boundary}$. The boundary is defined over $-0.506\leq(G_{\rm BP}-G_{\rm RP})_0\leq3.331$; no extrapolation is applied outside this interval.}
\label{fig:S1}
\end{figure}

\clearpage

\begin{table}[p]
\centering
\caption{\textbf{Spectroscopic validation of the fiducial massive-star selection.}}
\label{tab:S1}
\vspace{0.5em}
\small
\begin{tabular}{llrrr}
\toprule
Reference sample & Spectral subset & Benchmark $N$ & Recovered & Recovery fraction \\
\midrule
GOSC & All eligible entries & 516 & 228 & 44.2\% \\
\addlinespace
ALS~III G3 & All G3 sources & 2,158 & 757 & 35.1\% \\
ALS~III G3 & Secure O--B2 & 1,496 & 584 & 39.0\% \\
ALS~III G3 & O & 697 & 264 & 37.9\% \\
ALS~III G3 & B0 & 298 & 131 & 44.0\% \\
ALS~III G3 & B1 & 287 & 123 & 42.9\% \\
ALS~III G3 & B2 & 214 & 66 & 30.8\% \\
\bottomrule
\end{tabular}

\vspace{0.75em}
\begin{minipage}{0.94\textwidth}
\footnotesize
For GOSC, the benchmark comprises catalogue entries with at least one Gaia DR3 counterpart in the parent sample; an entry is counted as recovered when at least one matched Gaia source satisfies the fiducial selection. For ALS~III, the comparison uses Gaia DR3 source identifiers directly. Recovery fractions are conditional on inclusion in the parent sample and should not be interpreted as estimates of the completeness or purity of the photometrically selected sample. For comparison, a simple selection requiring $T_{\rm eff,GSP\mbox{-}Phot}>10,000~{\rm K}$ recovered 191 of the same 516 eligible GOSC entries (37.0\%), compared with 228 (44.2\%) recovered by the fiducial photometric selection.
\end{minipage}
\end{table}